\documentclass[usenatbib]{elsarticle}
\usepackage[english]{babel}
  \newcommand{\exclude}[1]{}
\usepackage[utf8x]{inputenc}
\usepackage{amsfonts}
\usepackage{amsmath}
\usepackage{amssymb}
\usepackage{xcolor}
\usepackage{enumitem}
\usepackage{graphicx}
\usepackage{gensymb}
\usepackage{fancyhdr}
\usepackage{times}
\usepackage{ulem}
\usepackage[caption = false]{subfig}
\usepackage{lineno}
 
\def\eV{\rm eV}
\def\<{\langle}
\def\>{\rangle}
\renewcommand{\d}{{\rm d}}
\def\+{\dagger}

\def\U1A{U(1)$_{\rm A}$}

\def\ra{\rangle}
\def\la{\langle}
\def\rmd{\mathrm{d}}

\newcommand{\be}{\begin{eqnarray}}
\newcommand{\ee}{\end{eqnarray}}
\newcommand{\beq}{\begin{equation}}
\newcommand{\eeq}{\end{equation}}

\begin{document}
\begin{frontmatter}
\title{Infrasonic, acoustic and seismic waves produced by the Axion Quark Nuggets}
 
\author[1]{Dmitry Budker}
       \ead{budker@uni-mainz.de}
        \address[1]{Johannes Gutenberg-Universit{\"a}t Mainz - Helmholtz-Institut, GSI Helmholtzzentrum f{\"u}r Schwerionenforschung, 55128 Mainz,   Germany\\
        Department of Physics,
University of California, Berkeley,  CA, 94720-7300, USA}
\author[2]{Victor V. Flambaum}
       \ead{v.flambaum@unsw.edu.au}
        \address[2]{School of Physics, University of New South Wales, Sydney 2052, Australia\\
        Johannes Gutenberg-Universit{\"a}tMainz (JGU) - Helmholtz-Institut, GSI Helmholtzzentrum f{\"u}r Schwerionenforschung 55128 Mainz,   Germany}

 \author[3]{Ariel Zhitnitsky}
\ead{arz@phas.ubc.ca}
\address[3]{
 Department of Physics and Astronomy, University of British Columbia, Vancouver, Canada}
 
\label{firstpage}

\begin{abstract}
We   advocate an idea that   the Axion Quark Nuggets (AQN)     hitting   the Earth  can be detected  by analysing the infrasound, acoustic  and seismic waves which always  accompany the AQN's  passage in the  atmosphere and   underground.  
Our estimates for the infrasonic frequency
 $\nu\simeq 5$\,Hz and overpressure $\delta p\sim 0.3$\,Pa for relatively  large size dark matter (DM) nuggets  suggest  that sensitivity of presently available instruments  is already sufficient to detect   very  intense (but  very rare) events   today with existing technology. A study of much more frequent but less intense events requires a new type of instruments. We propose a detection strategy for a systematic  study to search for  such  relatively weak and frequent  events     by using Distributed Acoustic Sensing and briefly mention other possible detection methods.  
 
\exclude{
 We advocate an idea that some mysterious explosions,
 the so-called skyquakes, which have been known  for centuries    could be   a manifestation of the dark matter Axion Quark Nuggets (AQN) when they propagate  in the Earth's atmosphere.  We specifically study the  event which occurred on July 31-st 2008 and was properly recorded by the dedicated Elginfield Infrasound Array (ELFO) 
  and seismic stations
 near London, Ontario, Canada.   The infrasound detection was accompanied by non-observation of any meteors by an all-sky camera network. Our interpretation is based on  We propose a detection strategy for a systematic  study to search for  such  explosions originating from AQNs  by using Distributed Acoustic Sensing and briefly mention other possible detection methods.
Specific signals from AQN tracks may also be detected with an existing network of seismic stations.
}
\end{abstract}
\begin{keyword}
Axion, Dark Matter,   
\end{keyword}

\end{frontmatter}

\vspace{0.1in}



\section{Introduction and Motivation} 
 \exclude{In this work we discuss two naively unrelated stories. The  first one is  the study  of  a specific dark matter (DM) model, the  so-called axion quark nugget (AQN) model  \cite{Zhitnitsky:2002qa}, see a brief overview of this model below. The second one deals with plentiful observational evidence of mysterious explosions, skyquakes, manifested in the form of sound and infrasound, the nature of which remains unknown in spite of the long history of observations, with records going back over 200 years. In this work we present arguments suggesting that 
 skyquakes could be a manifestation of the dark matter AQNs 
 traveling in the atmosphere. 
 
 Skyquakes \cite{skyquakes} are unexplained acoustic events that sound like a cannon shot or a sonic boom coming from the sky, see for example, a description in a TV interview by a meteorologist 
 \cite{skyquakes-meteorologist}. The main message of this short interview is  that skyquakes could not be due to any seismic events or meteor-type events which are routinely recorded around the globe. These events could not be identified with any human activities which are also recorded. One should also add that similar events have been recorded  for centuries in different countries with different  environmental features. These events cannot be explained by military aircraft as  records of skyquakes appeared long before supersonic flights. 
 
Unfortunately, it is next to impossible to extract any useful quantitative information from the numerous but random and unsystematic records. Luckily, one such event which occurred on July 31-st 2008 was properly recorded by the dedicated Elginfield Infrasound Array (ELFO) near London, Ontario, Canada  \cite{ELFO}.   The infrasound detection was accompanied by non-observation of any meteors by an all-sky camera network, ruling out a conventional meteor source. Anthropogenic sources such as operations at the nearby Bruce Nuclear Power Plant or the  Goderich salt mine were also eliminated; a  local airport radar reported no aircraft in the area at the time. In addition to infrasound, impulses were also observed seismically as ground-coupled acoustic waves around Southwestern Ontario and Northern Michigan. This event is further discussed in detail in Sec.\,\ref{observations} where we analyze  the energetics, infrasound-frequency properties, and other characteristics 
 of the event and compare them with our estimates based on the AQN model. 
}
The main goal of the present work is to present a new   possible detection  methods
of the axion quark nuggets (AQN) when they are propagating in the Earth's atmosphere and underground.   
 The AQN dark matter  model  \cite{Zhitnitsky:2002qa} was invented long ago with a motivation to explain the observed  similarity between the dark matter and the visible densities in the Universe, i.e. $\Omega_{\rm DM}\sim \Omega_{\rm visible}$. 
The idea that dark matter may take the form of composite objects of 
standard-model quarks in a novel phase goes back to quark nuggets  \cite{Witten:1984rs}, strangelets \cite{Farhi:1984qu}, and nuclearities \cite{DeRujula:1984axn},  see also the review \cite{Madsen:1998uh} with references to the original results. 
In the models \cite{Witten:1984rs,Farhi:1984qu,DeRujula:1984axn,Madsen:1998uh},  the presence of strange quark stabilizes the quark matter at sufficiently 
high densities allowing strangelets formed in the early universe to remain stable 
over cosmological timescales. 
 This type of DM  is ``cosmologically dark'' not because of the weakness of the AQN interactions, but due to their small cross-section-to-mass ratio, which scales down many observable consequences of an otherwise strongly-interacting DM candidate. We review the basics ideas, predictions and consequences  of this model in Sec.\,\ref{AQN-flux} and mention the crucial ingredients here. 

There are several additional elements in the AQN model in comparison with the older well-known and well-studied constructions \cite{Witten:1984rs,Farhi:1984qu,DeRujula:1984axn,Madsen:1998uh}. First, there is an additional stabilization factor for the nuggets provided by the axion domain walls which   are copiously produced  during the  quantum chromodynamic (QCD)  transition, which help to alleviate a number of  problems with the original \cite{Witten:1984rs,Farhi:1984qu,DeRujula:1984axn,Madsen:1998uh} nugget model\footnote{\label{first-order}In particular, a first-order phase transition is not a required feature for nugget formation as the axion domain wall (with internal QCD substructure)  plays the role of the squeezer. Another problem with \cite{Witten:1984rs,Farhi:1984qu,DeRujula:1984axn,Madsen:1998uh} is that nuggets likely evaporate on the Hubble time-scale. For the AQN model, this is not the case because the vacuum-ground-state energies inside (the color-superconducting phase) and outside the nugget (the hadronic phase) are drastically different. Therefore, these two systems can coexist only in the presence of an external pressure, provided by the axion domain wall. This should  be contrasted with the original model \cite{Witten:1984rs,Farhi:1984qu,DeRujula:1984axn,Madsen:1998uh}, which must provide stability at zero external pressure.}.  
Another feature of AQN which plays
a crucial role in the present work  is that nuggets can be made of {\it matter} as well as {\it antimatter} during the QCD transition. The direct consequence of this feature, given that the total baryon charge of the Universe is zero, is that DM density, $\Omega_{\rm DM}$, and the baryonic matter density, $\Omega_{\rm visible}$, will automatically assume the  same order of magnitude  $\Omega_{\rm DM}\sim \Omega_{\rm visible}$ without any fine tuning.

One should emphasize that AQNs are absolutely stable configurations on cosmological scales. The antimatter which is hidden in the form of the dense nuggets is unavailable for annihilation unless the AQNs hit stars or planets. There are also rare events of annihilation in the center of the galaxy, which, in fact, may explain some observed galactic excess emissions in different frequency bands, see Sec.\,\ref{AQN-flux}.  
\exclude{This is because they have the same QCD origin  and are both proportional to the same fundamental dimensional parameter $\Lambda_{\rm QCD}$  which ensures that the relation $\Omega_{\rm DM}\sim \Omega_{\rm visible}$  
always holds irrespective of the parameters of the model. 
}
The AQNs composed of antimatter are capable of releasing a significant amount of  energy when they enter the atmosphere and annihilation processes occur between  antimatter in the AQNs and atmospheric material. 
\exclude{
In the present work, we argue that skyquakes can be identified with AQN-annihilation events which inevitably occur when AQNs propagate in the atmosphere. 
We support this identification by estimating  the released energy, radiation characteristics, and the geometric and temporal evolution.
All these characteristics are consistent with the identification:
\be
\label{identification}
{\rm [skyquakes] } ~~\equiv ~~~ {\rm [AQN ~ annihilation~ events].}
\ee
}

How can this enormous amount of energy manifest itself? What would be the best option (optimal path) to detect the corresponding effects due to the AQN-annihilation processes? The problem is that the weakly interacting particles (such as the axions and neutrinos)
are hard to detect due to their weak interactions,
while relatively strongly interacting photons and leptons have short mean free path such that it is hard  to recover the origin of these particles. 

Indeed, 
the corresponding 
emission of photons in dilute environments such as the galactic center and at the locations of high-altitude Earth orbits
can be studied and properly analyzed  because the mean free path in such dilute environments is long. The corresponding results are discussed in Sec.\,\ref{AQN-flux}. This  should be contrasted with the case of dense environments where the  annihilation events  occur in the   Earth atmosphere when   the energy is released    in the form of the weakly coupled axions,  neutrinos as well as x and $\gamma$ rays and small number of   low-energy electrons and positrons.  It is hard to observe axions and neutrinos due to their feeble interactions, though the corresponding computations have been carried out recently, see Sec.\,\ref{AQN-flux}. At the same time, the  x and $\gamma$ rays emitted by AQNs are absorbed over short distances $\sim 10$\,m or so in the atmosphere, and therefore cannot be easily recovered for analysis due to the background radiation\footnote{The low-energy free electrons strongly interact  with  atmospheric material  while the low energy  positrons will be quickly annihilated such that it is hard to trace the origin of these particles emitted by AQNs.}. 

We propose in this work that  the AQN-induced signal can be studied by analysing the acoustic waves  which may propagate over large distances due to long absorption lengths for such waves. Furthermore, we shall argue that the corresponding signal   can be discriminated from background noise. Therefore, in this work  we describe    a new detection strategy of  the AQN based on their acoustic and seismic   manifestations.  

 \exclude{
We should emphasize that the AQN model was not designed to  explain the skyquakes. Rather, the AQN model
was invented for completely different purposes, to explain the observed relation: $\Omega_{\rm DM}\sim \Omega_{\rm visible}$. This model also predicts that there must be some acoustic shock waves as a result of AQN propagating in atmosphere and underground. We identify these acoustic shock waves with the skyquakes.

 Note that radiation from the AQN annihilation comes mainly in the form of X-rays with very small portion of the  visible light due to the secondary processes in atmosphere. 
 
This should be contrasted with conventional meteors which  typically emit visible light  due to their heating resulting from  interaction meteors with the atmosphere.  Visible light from conventional meteors is  routinely recorded using  all-sky cameras around the globe and represents an important element in identification and analysis of the events.  

 }
The presentation is organized as follows. In Sec.\,\ref{AQN-flux} we review the AQN model in the context of the present  work, paying special attention to the size distribution, frequency of appearance,  and the energy-emission pattern. In Sec.\,\ref{estimates}, we present our estimates for the AQNs propagating in the atmosphere and underground. The main lesson of this analysis is that the presently available instruments are capable to detect the signal produced by the large-size and intense AQNs which occur only once every 10 years or so. However, the presently available technical tools are not sufficiently sensitive
to study typical and relatively small AQNs. 
Therefore, 
 in Sec.\,\ref{detection-strategy}, we present a proposal for a systematic  study of acoustic and seismic events originating from AQNs with relatively small typical size which occur approximately once a day. We propose to 
   employ  Distributed Acoustic Sensing which uses
 optical-fiber  cables.  
 In  \ref{observations} as a side note
we  describe the observed mysterious  event which occurred on July 31st 2008 and was properly recorded by the dedicated Elginfield Infrasound Array (ELFO) \cite{ELFO}, near London, Ontario, Canada.   The infrasound detection was accompanied by non-observation of any meteors by an all-sky camera network. The signal was correlated with 
   seismic signals in the area. It is  tempting to identify the mysterious event recorded  by ELFO with a very intense  (and very rare)  AQN annihilation event which has precisely the required  features as described in Sec.\,\ref{estimates}.

\section{AQN model: the basics   }\label{AQN-flux}

The main goal of this section is to review the basic ideas 
of the AQN model, its motivation, consequences, and (as of yet) indirect, rather than direct,    supporting observations.  

The original motivation for the model can be explained as follows. 
It is commonly  assumed that the Universe 
began in a symmetric state with zero global baryonic charge 
and later (through some baryon-number-violating process, nonequilibrium dynamics, and $\cal{CP}$-violation effects, realizing the three  famous  Sakharov criteria) 
evolved into a state with a net positive baryon number.

As an 
alternative to this scenario we advocate a model in which 
``baryogenesis'' is actually a charge-separation (rather than charge-generation) process 
in which the global baryon number of the universe remains 
zero at all times.  In this model, the unobserved antibaryons come to comprise 
the dark matter in the form of dense nuggets of antiquarks and gluons in the  color superconducting (CS) phase.  
The result of this ``charge-separation process'' is two populations of AQN carrying positive and 
negative baryon number. In other words,  the AQN may be formed of either {\it matter or antimatter}. 
However, due to the global  $\cal CP$ violating processes associated with the so-called initial misalignment angle $\theta_0$ which was present  during 
the early formation stage,  the number of nuggets and antinuggets 
  will be different\footnote{The $\theta_0\neq 0$ is the source of strong ${\cal CP}$ violation in QCD. It represents a fundamental coupling
  constant of the theory, which  at the present epoch is experimentally observed to be very small, $\theta_0\lesssim 10^{-10}$.   Explaining this smallness represents the celebrated strong ${\cal CP}$ problem.
The most compelling resolution of the strong ${\cal CP}$ problem is formulated in terms of the dynamical   axion, such that an arbitrary $\theta_0\neq 0$ (at the QCD epoch during the evolution of the Universe)  relaxes to zero at present time,  see the original axion papers \cite{axion,KSVZ,DFSZ}, and   recent reviews \cite{vanBibber:2006rb, Asztalos:2006kz,Sikivie:2008,Raffelt:2006cw,Sikivie:2009fv,Rosenberg:2015kxa,Marsh:2015xka,Graham:2015ouw,Ringwald:2016yge,Irastorza:2018dyq}.}.
 This difference is always an order-of-one effect irrespective of the 
parameters of the theory, the axion mass $m_a$ or the initial misalignment angle $\theta_0$.
We refer to the original papers   \cite{Liang:2016tqc,Ge:2017ttc,Ge:2017idw,Ge:2019voa} devoted to the specific questions  related to the nugget formation, generation of the baryon asymmetry, and the 
survival   pattern of the nuggets during the evolution in  early Universe with its unfriendly environment.  

It is known  that the galactic spectrum 
contains several excesses of diffuse emission, the origin of which is not well established and  remains to be debated.
The best-known example is the strong galactic 511~keV line. If the nuggets have a baryon 
number in the $\langle B\rangle \sim 10^{25}$ range they could offer a 
potential explanation for several of 
these diffuse components. It is a nontrivial consistency check that the  $\langle B\rangle$ required to explain these excesses of the galactic diffuse emission belongs to the same mass range as discussed below. 
 For further details, see the original works \cite{Oaknin:2004mn, Zhitnitsky:2006tu,Forbes:2006ba, Lawson:2007kp,Forbes:2008uf,Forbes:2009wg} with explicit  computations 
 of the galactic-radiation  excesses  for various frequencies, including the excesses of the diffuse  x and   $\gamma$ rays.  
In all these cases, photon emission originates 
from the outer layer of the nuggets known as the electrosphere, and all intensities in different frequency bands are expressed in terms of a single parameter $\langle B\rangle$ such that all relative intensities   are unambiguously fixed because  they are determined by the Standard-Model (SM) physics.  

The  AQNs may also offer a resolution to the so-called ``Primordial Lithium Puzzle" \cite{Flambaum:2018ohm} and the ``Solar Corona Mystery" where the corona heating is explained by  the AQN annihilation events as argued in \cite{Zhitnitsky:2017rop,Raza:2018gpb}. Furthermore, the corresponding annihilation events are accompanied by the emission in radiofrequency bands which might  already have been observed as advocated in \cite{Ge:2020xvf}.  Furthermore, the same model with the same set of parameters  may resolve \cite{Zhitnitsky:2019tbh} the longstanding puzzle with the DAMA/LIBRA  observation  of the annual modulation at a $9.5\sigma$ confidence level, which is in direct conflict with other DM experiments if interpreted in terms of the WIMP interaction with nuclei. Finally, the annihilation events when an AQN traverses the Earth atmosphere under the thunderstorm (where strong electric fields are present) may mimic the ultrahigh-energy cosmic rays. The recent unusual events recorded by  Telescope Array Experiment might be precisely those types of events  as argued in \cite{Zhitnitsky:2020shd}.

The key parameter which essentially determines all the intensities for the effects mentioned above is the average baryon charge $\la B \ra$ of the AQNs. There are a number of constraints on this parameter which are reviewed below. 
One should also mention that the AQN mass is related to its baryon charge by $M_N\simeq m_p |B|$, where we ignore small differences between the energy per baryon charge in  CS   and  hadronic confined  phases.  The AQNs are macroscopically large objects with a typical size of $R\simeq 10^{-5}{\rm cm}$ and roughly nuclear density of order $10^{40}\,{\rm cm^{-3}}$  resulting in masses  of roughly 10\,g.  

An event where an AQN impinges on the Earth should be contrasted with conventional meteors. A conventional object  with mass 10\,g would have a typical size of order 1\,cm occupying the volume which would be 15 orders of magnitude larger than the AQN volume. This is due to the  fact that AQNs have nuclear density which is 15 orders of magnitude higher than   the density of normal matter.   One can view an AQN as a small neutron star (NS) with its nuclear density. The difference is that  a NS is squeezed by gravity, while an AQN is squeezed by the axion-domain-wall pressure. The drastic density difference between AQNs and conventional small meteors leads to fundamentally   different  interaction patterns  when they enter the Earth atmosphere.  

We now turn to the observational constraints on such kind of dense objects. 
The strongest direct-detection limit 
is  set by the IceCube Observatory,  see Appendix A in \cite{Lawson:2019cvy}:
\be
\label{direct}
\la B \ra > 3\cdot 10^{24} ~~~[{\rm direct ~ (non)detection ~constraint]}.
\ee
The authors of \cite{Herrin:2005kb} used the Apollo data to constrain the abundance of quark nuggets in the range of 10\,kg to one ton. They  argued that the contribution of such heavy nuggets  must be at least an order of magnitude less than would saturate the dark matter in the solar neighbourhood \cite{Herrin:2005kb}. Assuming that the AQNs do saturate the dark matter, the constraint  \cite{Herrin:2005kb} can be reinterpreted as at least $90\%$ of the AQNs having masses below 10\,kg. This constraint can be expressed  in terms of the baryon charge:
   \be
\label{apollo}
\la B \ra \lesssim   10^{28} ~~~ [  {\rm   Apollo~ constraint ~} ].
\ee
Therefore, indirect observational constraints (\ref{direct}) and (\ref{apollo}) suggest that if the AQNs exist and saturate the dark matter density today, the dominant portion of them   must reside in the window: 
\be
\label{window}
3\cdot 10^{24}\lesssim\la B \ra \lesssim   10^{28}~ [{\rm constraints~ from~ observations}].  ~~~
\ee

We emphasize that the AQN model with the limits (\ref{window})  is consistent with all presently available cosmological, astrophysical, satellite and ground-based constraints. This model is rigid and predictive since there is little flexibility or freedom to modify any estimates mentioned above 
In particular, the AQN flux (\ref{eq:D Nflux 3}) which plays a key role in the present studies cannot  change by  more than a factor of two, depending on the size distribution within the window 
(\ref{window}).    

 It is important  that the frequency of appearance of AQNs depends on  the size distribution  $f(B)$ defined as follows:
 Let $\rmd N/\rmd B$ be the number of AQNs which carry the baryon charge [$B$, $B+dB$].
The mean value of  the baryon charge $\langle B\rangle $ is given by 
\begin{equation}
\label{eq:f(B)}
\langle B\rangle 
=\int_{B_{\rm min}}^{B_{\rm max} }\rmd B~B f(B), 
~~~~~ f(B)\propto B^{-\alpha}, 
\end{equation}
where $f(B)$ is a  properly normalized distribution  and $\alpha\simeq (2-2.5)$  is the power-law index.
One should emphasize that the parametrization  (\ref{eq:f(B)}) was suggested in solar physics studies  to fit the observed extreme UV radiation from the entire solar surface. We adopted this scaling in \cite{Zhitnitsky:2017rop,Raza:2018gpb}, where it was proposed   that  the so-called  nanoflares (conjectured by Parker many years ago to resolve the ``Solar Corona Mystery")  can be  identified with AQN-annihilation events in the solar corona. The main motivation for this identification  is that  the observed intensity of the extreme UV emission from the solar corona matches the total energy released as a result of   the AQN-annihilation events in the transition region assuming the conventional value for the dark matter density around the Sun, $\rho_{\rm DM}\simeq 0.3\,{\rm  {GeV} {cm^{-3}}}$.  One should emphasize that this ``numerical coincidence"  is a highly nontrivial self-consistency check of the  proposal \cite{Zhitnitsky:2017rop,Raza:2018gpb} connecting nanoflares with AQNs, since the nanoflare properties  are  constrained by solar corona-heating models, while the intensity of the extreme UV due to the AQN annihilation events  is mostly determined by the dark matter density\footnote{Furthermore,  the required energy interval for the    nanoflares  must be  in the range: $ E_{\rm nano}   \simeq     ( 10^{21}  -  10^{26})~{\rm erg}$. This allowed interval   largely overlaps with the AQN baryonic charge window (\ref{window}) if the identification between nanoflares and AQN annihilation events is made. In this case, 
 $E_{\rm nano}\simeq 2 m_pc^2 B \ \simeq (3\cdot 10^{-3} \ {\rm erg}) \cdot B$,  see \cite{Zhitnitsky:2017rop,Raza:2018gpb} for details.}.
 One should  note that the algebraic scaling (\ref{eq:f(B)}) is a generic feature of the AQN-formation mechanism based on percolation theory  \cite{Ge:2019voa},   but $\alpha$  cannot be theoretically computed in strongly coupled QCD. 

 We now estimate the rate at which AQNs hit the Earth assuming the local dark matter density of $\rho_{\rm DM}\simeq 0.3\,{\rm  {GeV} {cm^{-3}}}$. Assuming the conventional halo model one arrives at \cite{Lawson:2019cvy}:
 \be
\label{eq:D Nflux 3}
\frac{\langle\dot{N}\rangle}{4\pi R_\oplus^2}
&\simeq & \frac{4\cdot 10^{-2}}{\rm km^{2}~yr}
\left(\frac{\rho_{\rm DM}}{0.3{\rm \frac{GeV}{cm^3}}}\right)
 \left(\frac{\langle v_{\rm AQN} \rangle }{220~{\rm \frac{km}{s}}}\right) \left(\frac{10^{25}}{\langle B\rangle}\right),\nonumber \\
 \langle\dot{N}\rangle&\simeq & 0.67\,{\rm s}^{-1} \left(\frac{10^{25}}{\langle B\rangle}\right)\simeq  2.1\cdot 10^7 {\rm yr}^{-1} 
\left(\frac{10^{25}}{\langle B\rangle}\right).
\ee
Averaging over all types of AQN trajectories  with different masses $ M_N\simeq m_p|B|$,  different incident angles, different initial velocities and size distributions does not significantly modify this estimate.   The result (\ref{eq:D Nflux 3}) suggests that 
the AQNs hit the Earth surface with a frequency  approximately  once a day   per   $100\times100\,{\rm km^2}$ area.    This rate is suppressed for large AQNs according to the distribution function (\ref{eq:f(B)}).
 
 It is instructive to compare the rate (\ref{eq:D Nflux 3}) with  the  number of   meteoroids which enter  the Earth atmosphere. This rate is of order $10^8 {\rm day^{-1}}$, see review \cite{Silber-review}.        It is more informative to represent this rate in terms of the total mass of the falling meteoroids, which is
 $10^5$\,tons/year and is much greater  than the total mass of order $5\cdot 10^2\,{\rm tons/year}$ associated with the dark matter AQN rate (\ref{eq:D Nflux 3}). The size distribution for meteoroids peaks  at $2\cdot 10^{-2}\,{\rm cm}$ while the mass distribution peaks at around $10\,{\rm \mu g}$, see  \cite{Silber-review}. This should be contrasted with a typical AQN size of $R\simeq 10^{-5}{\rm cm}$ and  mass of   roughly 10\,g. 
 
 The final topic to be reviewed here is the spectral  properties  of the AQN emission.    The most important feature of the  AQN spectrum  which  distinguishes it  from meteor emission is that the AQN spectrum is peeked in the (10-50)\,keV range, while the optically visible bands at $\sim (1-10)\,$eV are strongly suppressed, see  Fig.\,\ref{spectrum} in   \ref{sect:spectrum}. Another crucial difference with meteor emission is that the  AQN spectrum is not thermal black body radiation as it originates from the annihilation events. It should be contrasted with the black body radiation  of  conventional meteors and meteorites   entering the Earth atmosphere with supersonic velocities and experiencing friction with  the surrounding material resulting in    heating  of the   meteoroids and surrounding material. We refer to   \ref{sect:spectrum} discussing the spectral features of the AQNs traversing the atmosphere, see in particular Fig.\,\ref{spectrum}.

 \exclude{
 These features in the spectrum have  some profound consequences for the present work because  AQNs do not emit significant amounts of visible light, in contrast with conventional meteors   which are normally characterized by  strong emission in the visible frequency bands through sputtering and ablation \cite{Silber-review, Silber-optical-1, Silber-optical-2}. \textbf{DB: I am not sure what emission thorough sputtering and  ablation means exactly; it does not sound right to me (I am a guy who actually uses sputtering and ablation in the lab); also everything in this paragraph is a duplication of the information given elsewhere; I propose to cut this whole paragraph.}
 }
 These features in the spectrum    imply  that the AQNs cannot be observed by conventional optical monitoring  as AQNs are not   accompanied by significant emission of the visible light and cannot be routinely observed by all-sky cameras.    Therefore, the observation of a signal by  infrasound instruments  and   non-observation by the   optical   synchronized cameras (which must  continuously  monitor   the sky by recording the conventional meteors) would unambiguously 
 identify the AQNs entering the atmosphere. This  synchronization   eliminates or dramatically reduces all   possible  spurious events
 from sky. 
 
   \section{  Acoustic signals from   meteoroids and AQNs }\label{estimates}
   \subsection{Blast wave from meteoroids} 
     We start by reviewing a  
     \exclude{conventional framework  of  the}  model \exclude{which was} designed to study meteor-generated infrasound  \cite{ReVelle} 
   (originally this model was introduced to describe a blast wave from a lightning discharge, so it has a general character).
   There were many recent advances \exclude{and  improvements} in this framework including the  comparison with observational data   \cite{Silber-optical-1,Silber-optical-2}. Our goal here is to use this framework    to \exclude{apply the basic ideas  of Ref.\,\cite{ReVelle} to} estimate   the intensity and frequency characteristics of the infrasound signal generated by AQNs   propagating in the atmosphere.  Our estimates cannot literally follow  \cite{ReVelle,Silber-optical-1,Silber-optical-2} as the nature of the released energy in the case of AQN is drastically different from the energy sources associated with    conventional meteors 
   in the Earth atmosphere. 
   \exclude{\textbf{DB: the following text is the N-th repetition of the same thought...Propose deleting: In the former case the source is the annihilation energy between atoms and molecules from the atmosphere with antimatter hidden in the AQNs, while in  the later case, the energy source  is friction of the meteors with  the surrounding material  leading to heating   and radiation.} 
   }
   However, we think that the  generic scaling features describing the sound waves at large distances hold in both cases. \exclude {which is precisely the key element  to be used for the estimates which follow.} Furthermore, the Mach number $M=v/c_s\gg 1$ (here $v$ is the speed of the meteor and $c_s$ is the speed of sound) is very large for meteors as well as for AQNs such that cylindrical symmetry is assumed to hold for propagating sound and infrasound waves  in both cases. 
   
   The basic parameter  of the approach  \cite{ReVelle,Silber-optical-1,Silber-optical-2} is the so-called characteristic blast-wave relaxation radius defined as 
   \be
   \label{R_0}
   R_0\equiv \sqrt{ \frac{E_l}{p_0}  },
   \ee
   where $E_l$ is the energy deposited by the meteor per unit trail length, and $p_0$ is the hydrostatic atmospheric pressure. 
   The physical meaning of this parameter $R_0$ is the distance at which the overpressure approximately equals the hydrostatic atmospheric pressure. In the case of a bomb-like explosion, the relevant parameter can be defined as 
    \be
   \label{R_1}
   R_1\equiv \sqrt[3]{ \frac{E_{\rm point~ source}}{p_0}  },
   \ee
   where $E_{\rm point~ source}$ is the energy deposited to the air as a result of   explosion. The parameter $R_1$ has the same physical meaning as $R_0$ and it   determines  the distance at which the overpressure approximately equals to the hydrostatic atmospheric pressure.
   
   In simple cases for  meteors, the parameter $R_0$ can be directly expressed in terms of the Mach number $M$ and the meteor diameter $d_m$ as $R_0\sim M d_m$, see \cite{ReVelle,Silber-optical-1,Silber-optical-2}. The significance of the parameter $R_0$ is that the overpressure $\delta p$ at larger distances can be expressed in terms of dimensionless parameter $x$ defined as follows \cite{ReVelle,Silber-optical-1,Silber-optical-2}:
   \be
   \label{delta_p}
   \frac{\delta p}{p_0}=\frac{2(\gamma+1)}{\gamma} f(x), ~~~ x\equiv \frac{r}{R_0}, ~~~ f(x\gg 1) \simeq  \frac{1}{x^{3/4}},~~
   \ee
   where $\gamma=c_p/c_v$. Note that the overpressure $\delta p$ decays faster than $r^{-1/2}$ as it would be for a cylindrical sound wave with a given frequency. This is due  to increase of the  width $l$ of the blast wave packet as follows:  $l \sim R_0 x^{1/4}$. Correspondingly, the fundamental sound frequency $\nu$  decreases as $\nu \sim c_s/l \sim (c_s/R_0)x^{-1/4}$, where $c_s$ is the speed of sound.
   
   The scaling \eqref{delta_p} is justified when overpressure is relatively small. 
   In case of conventional meteors, all parameters such as $R_0$ can be modelled and compared with observations  \cite{Silber-optical-1,Silber-optical-2}.
   We do not have such luxury in the AQN studies. However, some theoretical estimates can be made, which is \exclude{precisely} the topic of this section.
   
   \subsection{AQN in the atmosphere}\label{atmosphere}
   We start with estimation of the parameter $E_l$ entering (\ref{R_0}). In the AQN framework, the energy of annihilation events occurring per unit length while the nugget traverses the atmosphere is: 
   \be
\label{E_l}
 E_l &\simeq &  \kappa \cdot (\pi R^2) \cdot (2~ {\rm GeV})\cdot n_{\rm air} \\
 &\simeq& 10^4\cdot \kappa  \left( \frac{B}{10^{25}}\right)^{2/3} \left(\frac{n_{\rm air}}{10^{21} ~{\rm cm^{-3}}}\right)\frac{\rm J}{\rm m},\nonumber
   \ee
   where  $n_{\rm air}$ is the total number of nucleons  in atoms such that $\rho =n_{\rm air} m_p$. The parameter $\kappa$ as explained in  \ref{sect:spectrum}   is introduced to account for the fact that not all matter striking the nugget will 
annihilate and not all of the energy released by an annihilation will be thermalized in the nuggets (for example, some portion of the energy will be released in the form of the axions and neutrinos), see the discussion after Eq.\,(\ref{eq:rad_balance}).
As such $\kappa$ encodes a large number of complex processes including the probability that 
not all atoms and molecules may be able to penetrate into the 
color superconducting phase of the nugget to get annihilated. It also includes complicated dynamics due to the very large Mach number $M=v_{\rm AQN}/c_s\gg 10^2$ when shock waves are formed and the turbulence has developed. Both phenomena lead to efficient energy exchange between the nugget and surrounding material.
Assuming a $10$ km   as a typical length scale where emission occurs one can estimate the total released energy in atmosphere  on the level $10^8 J$, which of course represents a tiny portion $\sim  10^{-7}$ of the total energy hidden in the AQN.  For simplicity   we keep $\kappa\simeq 1$ in our order of magnitude estimates which follow. 

Directly using the estimate (\ref{E_l}), one arrives at the following approximate expression for the parameter $R_0$:
\be
\label{R_AQN}
  R_0^{\rm AQN}\equiv \sqrt{ \frac{E_l}{p_0}  }\sim 0.3 ~ \left( \frac{B}{10^{25}}\right)^{\frac{1}{3}}\left({\frac{n_{\rm air}}{10^{21} ~{\rm cm^{-3}}}}\right)^{\frac{1}{2}}  {\rm m},~~~~
\ee
   which has a physical meaning of a distance where overpressure due to the AQN annihilation events equals the hydrostatic atmospheric pressure.
   
   Several comments are in order regarding this estimate. In the case of conventional 
   or nuclear explosion  the blast occurs as a result of the interaction of the radiation with surrounding material which rapidly heats the material. This causes vaporization of the material resulting in its rapid expansion, which eventually contributes to formation of the shock-wave. All these effects occurring in conventional 
   explosions 
     at very small scales, much smaller than a typical radius where over-pressure approximately equals to atmospheric pressure. In case of cylindrical symmetry the relevant parameter is determined by $R_0$ in Eq.\,(\ref{R_0}). In case of a point-like explosion the corresponding distance $R_1$ is determined by (\ref{R_1}) which plays the role of $R_0$ for point-like explosion.
     
     Now we estimate the distances where the radiation is effectively converted to the
     shock-wave energy. In case of conventional or nuclear explosion 
    the dominant portion of   the radiation comes  in the $ 20~{\rm eV}$ energy range and above. At this energy, the dominant process is the atomic photoelectric effect with cross section $\sigma_{p.e.}\sim 10^7 {\rm ~barn} $ and higher
   such that the photon   attenuation length  $\lambda \sim 10^{-6}{\rm g/cm^{2}}$, 
   see e.g. Fig.\,33.15 and Fig.\,33.16 in \cite{PDG} and references therein. 
   In case of meteoroids the emission  normally occurs in $(1-20)$ eV energy range, which also includes the visible light emission.
  These spectrum features in air imply  that the energy due to the heating will be completely absorbed on the scales which are much shorter than $R_0$ defined by (\ref{R_0}), i.e. 
   \be
   \label{lambda}
   \frac{\lambda}{\rho_{\rm air}} \lesssim 10^{-3}{\rm cm}\ll R_0 ~~~~~ [\rm   meteoroids]. ~~ 
   \ee
   
   This should be contrasted with the AQN case with a drastically different radiation spectrum with typical energy in the $\sim 40$ keV range as reviewed in 
   \ref{sect:spectrum}. In this case the   atomic photoelectric effect 
   which is still the dominant process
and the photon  attenuation length is $\lambda\sim  0.5\,{\rm g/cm^{2}}$
such that 
     \be
   \label{lambda_AQN}
   L=\frac{\lambda}{\rho_{\rm air}} \sim 5 ~{\rm m}\gg R_0^{\rm AQN} ~~~~~ [\rm AQN~ events]. ~~ 
   \ee
   These estimates suggest that only a small portion of the energy (\ref{E_l}) will be released in the form of a blast, while the rest of the energy will simply heat the surrounding material. 
   The attenuation length is even longer for higher-energy photons which saturate the total intensity for $T_{\rm AQN}\simeq 40$ keV, see 
   \ref{sect:spectrum}.
   
  We can do an estimate of overpressure in this case as follows. The annihilation energy $E_l z$ released on the track of the length $z$ is absorbed in the volume of the cylinder $V=\pi L^2 z$. The internal energy of a diatomic ideal gas (air) is given by $U=(5/2) PV$. This gives an estimate of overpressure inside  this volume $V$:
   \be
   \label{AirDeltaP0}
   \delta p_0 \approx \frac{E_l}{2.5 \pi L^2} ~~~~~~ [\rm AQN~ events].
   \ee  
   As explained above, outside this volume,  $\delta p$ decreases as $f(\bar{x})\approx \bar{x}^{-3/4}$, where we introduced the dimensionless parameter $\bar{x}=r/L$, which plays the same role as $x$ in meteoroids formula  (\ref{delta_p}):
   \be
   \label{AirDeltaP}
   \delta p(r)\approx  \frac{\delta p_0}{\bar{x}^{3/4}} \approx  \frac{E_l}{2.5 \pi L^2} \left(\frac{1}{\bar{x}}\right)^{3/4}~~~ [\rm AQN~ events].~~~~
   \ee  
  To illustrate the significance of the estimate (\ref{AirDeltaP}), we present an order of magnitude  numerical estimate for the overpressure at a distance $r$, with the annihilation energy given by Eq.\,(\ref{E_l}) and absorption of this  energy within the radius $L=5$ m as estimated by (\ref{lambda_AQN}): 
 \be
\label{AirDeltaPnumber}
 \delta p\approx 0.03 {\rm Pa} \left( \frac{B}{10^{25}}\right)^{2/3}\left( \frac{100\, {\rm km}}{r}\right)^{3/4}~[\rm AQN~ events].~~~~~
 \ee
 This estimate  shows that  a typical AQN generates very small overpressure $\delta p/p \sim 10^{-3}  $ even inside the absorption region $r\lesssim L$, which  should be contrasted with the meteoroid case (\ref{delta_p}) where $\delta p/p \approx   1$
 at $r\simeq R_0$. The difference is due to the large   length $L\simeq 5$\,m in comparison with the small absorption distance $\sim 10^{-3}$cm  for the  meteoroids (\ref{lambda}).  The temperature increase  in surrounding region $\delta T/T \sim \delta p/p \sim 10^{-3} \ll 1$ 
 is too small to produce visible thermal radiation around the AQN path.\footnote{This temperature should not be confused with the much higher internal AQN temperature $T_{\rm AQN}\sim 10$\,keV.}

      Another important   characteristic of the acoustic  waves produced by meteoroids is the scaling behaviour  
      of the so called line-source wave period $  \tau (x)$ at large distances. The scaling behaviour 
      can be expressed in terms of the same dimensionless parameter $x$  introduced above, and it is given by \cite{ReVelle,Silber-optical-1,Silber-optical-2}:
   \be
   \label{tau}
   \tau (x)\simeq  0.562\tau_0 {x^{1/4}},  ~~ \tau_0=2.81 \frac{R_0}{c_s} ~~[\rm   meteoroids], ~~~~~
   \ee
where $\tau_0$ is the so-called fundamental period where numerical  factors $2.81$ and $0.562$ in Eq.\,(\ref{tau})  have been  fitted from the observations \cite{Silber-optical-2}.  Equation (\ref{tau}) determines the frequency of the sound (infrasound) wave at a distant point $x$:
\be
\label{frequency}
\nu (x) \equiv   \tau^{-1} (x) \sim   \tau_0^{-1}{x^{-1/4}}, ~ x\gg1 [\rm   meteoroids].
\ee

The same scaling behaviour is expected to hold for the AQN case. However, the parameters for AQNs are different:
\be
\tau_0 \sim   \frac{L}{c_s}, ~~ \tau (\bar{x})\sim   \tau_0 \, \bar{x}^{1/4} , ~~ \nu_0\equiv \frac{1}{\tau_0} \sim  70~ {\rm Hz},  
\ee
where we ignore all numerical factors which in the case of meteoroids were  fixed by matching with observations, and obviously cannot be applied  to our present studies of the AQNs.  In this case we arrive at the following estimate for the frequency at a distance $r$:
\be
\label{frequencyAQN}
\nu (\bar{x}) 
\sim  \frac{\nu_0}{{\bar{x}^{1/4}}}\sim 6\, {\rm Hz} \left( \frac{100\, {\rm km}}{r}\right)^{1/4}  ~[\rm AQN~ events].~~
\ee
Thus, at a large distance from the AQN track in the air there will be emission of the  infrasound waves of  low frequency.
We will see below that  for the signal from an underground AQN track,  the overpressure and the frequency are both several orders of magnitude higher.

   \exclude{
   
   ---------------------------------------------------------------------------------------
   
   It is hard to compute the relevant portion contributing  to formation of the shockwave from the first principles\footnote{\label{xi}For example, a strong increase  of energy exchange between AQN and surrounding material could occur  due to the elastic collisions of electrons with very hot nugget's core. Free electrons are not normally present in neutral atmosphere. However, they may emerge  as  a result of  ionization processes and turbulence which always  accompany the AQN supersonic motion. These effects may drastically reduce the effective attenuation length ${\lambda}/{\rho_{\rm air}}$ given by Eq.\,(\ref{lambda_AQN}) to bring it closer to $R_0^{\rm AQN}$, strongly   affecting  parameter $\xi$ introduced by Eq.\,(\ref{blast}).}. Therefore, for the estimates which follow we opted to introduce an unknown parameter  $\xi$ to account for this complicated physics:
   \be
   \label{blast}
   E_{\rm blast}\equiv \xi \cdot E_l, ~~~ \xi\lesssim (10^{-3}-10^{-4}) 
   \ee  
  
   
   To accommodate this modification  in energetics for AQNs we simply rescale parameter $R_0^{\rm AQN}$ such that  the overpressure 
   scaling behaviour will be modified   as follows:
     \be
   \label{delta_p_AQN}
   \frac{\delta p}{p_0}\simeq  f(\bar{x}), ~~ \bar{x}\equiv \frac{r}{\sqrt{\xi}R_0^{\rm AQN}}, ~~ f( \bar{x}) \simeq  \frac{1}{\bar{x}^{3/4}}. 
   \ee
   }   
\exclude{
However, we should take all numerical parameters, of course, must be different.
Most important  (parametrical)  distinct feature from conventional meteoroid case (\ref{tau}) is that the dimensional parameter $\tau_0 $ for the AQN case is not just $R_0^{\rm AQN}$ but may assume any value between $R_0^{\rm AQN}$ and the photon   attenuation length (\ref{lambda_AQN})
 as explained above. To be more precise the fundamental period $\bar{\tau}_0$  and the  line source wave period $  \bar{\tau} (\bar{x})$ at large distances are determined by the following relations:
  \be
   \label{tau_AQN}
  \bar{\tau}  (\bar{x})\sim   \bar{\tau}_0 {\bar{x}^{1/4}},  ~~~~~~ \frac{R_0^{\rm AQN}}{c_s}\lesssim \bar{\tau}_0\lesssim  \frac{\lambda}{c_s\rho_{\rm air}},  
   \ee
   where $\bar{x}$ is defined by Eq.\,(\ref{delta_p_AQN}) as before. 
  The  frequency of the sound (infrasound) wave for the AQN events at very distant point $r$ is determined by the following scaling relation:
\be
\label{frequency}
\bar{\nu} (\bar{x}) \equiv   \bar{\tau}^{-1} (\bar{x}) \sim  \bar{\tau}^{-1}_0 {\bar{x}^{-1/4}}, ~~  ~~ [\rm   AQN ~events].~~~~
\ee
}

\subsection{AQN propagating underground }\label{underground}

One should emphasize that the infrasound waves originating from AQNs as estimated in Sec.\,\ref{atmosphere} will be always accompanied by sound waves emitted by the same AQNs  when the nuggets hit the Earth surface and continue to propagate in  the deep  underground.  The corresponding estimates of intensity and frequency of the sound emitted as a result of the annihilation events occurring underground are presented in this subsection. 
 
 The starting point is similar to (\ref{E_l})   which for underground rocks  assumes the form:
    \be
    \label{E_r}
E_l^{\downarrow} &\simeq &  \kappa \cdot (\pi R^2) \cdot (2~ {\rm GeV})\cdot n_{\rm rock} \\
 &\simeq& 10^7 \cdot \kappa      \left( \frac{B}{10^{25}}\right)^{2/3} \left(\frac{n_{\rm rock}}{10^{24} ~{\rm cm^{-3}}}\right) \frac{\rm J}{\rm m},\nonumber
   \ee
  where we use the notation $E_l^{\downarrow}$ for the energy produced by annihilation (some of which may remain in the AQN) to avoid confusion with the similar equation (\ref{E_l})  applied to the atmosphere, $n_{\rm rock}$ is the total number of nucleons  in atoms such that $\rho =n_{\rm rock} m_p$. 
  
   We introduce   an unknown parameter  $\xi^{\downarrow}$ which applies to the underground case (at sufficiently high density of the surrounding material) to account for the complicated physics   which describes the transfer of the AQN energy into the the surrounding material energy  denoted as $E^{\downarrow}_{\rm blast}$:
   \be
   \label{blast_r}
   E^{\downarrow}_{\rm blast}= 10^7\cdot \xi^{\downarrow}  \left( \frac{B}{10^{25}}\right)^{2/3} \left(\frac{n_{\rm rock}}{10^{24} ~{\rm cm^{-3}}}\right) \frac{\rm J}{\rm m}.
   \ee  
   There are several important new  elements  in comparison with the atmospheric case discussed in Sec.\,\ref{atmosphere}.
   First of all, the increase of the density of the surrounding material naively drastically increases  the released annihilation energy
   as Eq.\,(\ref{E_r}) suggests assuming that the coefficient $\kappa$ remains the same as in the atmospheric case,  Eq.\,(\ref{E_l}).
   However, it is expected   that this assumption is strongly violated.\footnote{\label{xi}The main reason for that is due to increase of the internal temperature $T_{\rm AQN}$ which consequently leads to strong ionization of the positrons from electrosphere. As a result of this ionization  the positron density of the electrosphere (which is responsible for the emissivity)   drastically decreases. It    suppress  the emissivity from the electrosphere  as Eq.\,(\ref{eq:P}) states. It should be contrasted with the BB radiation where emissivity scales as $\sim T^4$. In our case, the source of emission are the positrons from electrosphere, not BB radiation.} If one removes the  low-energy positrons from the electrosphere, the suppression factor could be $\xi^{\downarrow}\sim 10^{-2}$ and even much smaller.\footnote{\label{meanfield}Such simplified  procedure for  the estimate of $\xi^{\downarrow}$ by  complete removing the low energy states is not a proper way of computation because the positron's density will be adjusting when   $T_{\rm AQN}$ varies.  The consistent procedure would be a mean-field   computation of the positron density by imposing the proper boundary conditions relevant for nonzero temperature and non-zero charge, similar to $T_{\rm AQN}\approx 0$ computations carried out in \cite{Forbes:2008uf,Forbes:2009wg}.
   The corresponding computations have not been done yet, and we keep the parameter $\xi^{\downarrow}(T_{\rm AQN})$ as a phenomenological free parameter.}
   
   The second important effect which was ignored in the atmosphere  in Sec.\,\ref{atmosphere} is that $\delta p$ could be much larger underground   in comparison with the estimate of Eq.\,(\ref{AirDeltaPnumber}).  It    results in pushing material from the AQN path which effectively   decreases  the geometrical cross section  $\pi R^2$   assumed in (\ref{E_r}). This effect further suppresses the parameter $\xi^{\downarrow}$ entering (\ref{blast_r}).
   
   We cannot at the moment compute   $\xi^{\downarrow}$ from first principles as mentioned in footnotes \ref{xi} and \ref{meanfield}. Therefore, we keep it as a phenomenological unknown parameter which strongly depends on the environment, temperature $T_{\rm AQN}$ and many complex processes as mentioned above.
   
   Another important parameter is the absorption length $L^{\downarrow}(T_{AQN})$ for the energy emitted by AQN in underground (hence the $^{\downarrow}$ label), which also indirectly depends on the AQN internal temperature $T_{AQN}$. This is because the length $L^{\downarrow} $ strongly depends on the energy of the photons emitted by the AQNs, which is determined by the internal temperature $T_{AQN}$. For the photon energy $\sim 100$\,keV, an absorption length in silicon is about $L^{\downarrow}\simeq 2$ cm. However, it is an order of magnitude larger for 1 MeV photons. We  account for this uncertainty by introducing another unknown dimensionless parameter $\eta$ defined as follows: $L^{\downarrow}(T_{AQN}) =(2{\rm cm})\cdot\eta$.
   In terms of these unknown  parameters the 
 deposited energy per unit volume $\epsilon_{\rm blast}$ surrounding the AQN can be estimated as follows:
 \be
 \label{deposited}
  \epsilon_{\rm blast}\simeq 10^2\frac{\rm J}{\rm cm^3}  \cdot\left(\frac{\xi^{\downarrow} }{10^{-2}}\right)\cdot\left(\frac{1}{\eta^2 }\right), \nonumber
 \ee
  which leads to an instantaneous increase of temperature $\Delta T$ of the surrounding material:
  \be
 \label{temperature}
  \Delta T\simeq  30~ K \cdot\left(\frac{\xi^{\downarrow} }{10^{-2}}\right)\cdot\left(\frac{1}{\eta^2 }\right).  
 \ee
  We now in position to estimate the overpressure for the blast wave in two different approximations. First, we may 
estimate the overpressure as deposited energy per unit volume. This yields
  \be
 \label{delta_p_0}
 \delta p\simeq 10^7~ {\rm Pa} \cdot\left(\frac{\xi^{\downarrow} }{10^{-2}}\right)\cdot\left(\frac{1}{\eta^2 }\right)~ {\rm at} ~r\simeq L^{\downarrow} .  ~~~
 \ee
   Another  approximation is based on an increase of pressure due to the thermal expansion of solids. Relative thermal expansion of silicon is $\delta x/x$=2.6 $10^{-6}$/K, the Young modulus is 150\,GPa $\sim 10^{11}$\,Pa. This gives 
    the same order of magnitude as in the dimensional estimate (\ref{delta_p_0}).   
 
 Our next task is to estimate the amplitude of the wave at large distance $r$.  Using the conventional scaling arguments when $\delta p (r) \sim 1/(\bar{x}^{\downarrow})^{3/4}$ with dimensionless parameter $\bar{x}^{\downarrow}$ defined as   $\bar{x}^{\downarrow}=r/L^{\downarrow}$ we arrive to the following estimate for   overpressure at a distance $r$:
 \be
   \label{SolidDeltaP}
  \delta p(r) \sim 10^2\cdot\left(\frac{\xi^{\downarrow}}{10^{-2}}\right)\cdot\left(\frac{1}{\eta }\right)^{\frac{5}{4}} \cdot\left(\frac{100\,{\rm km}}{r}\right)^{\frac{3}{4}}~{\rm Pa}.~~~
   \ee 
 Following  the same logic as for Eq.\,(\ref{frequencyAQN}) we obtain a numerical estimate for the  frequency of sound emitted by an AQN propagating underground:
\be
\label{nu0}
\tau_0 \sim   \frac{L^{\downarrow}}{c_s}, ~~ \tau (\bar{x}^{\downarrow})\sim   \tau_0 \cdot \sqrt[4]{\bar{x}^{\downarrow}} , ~~ \nu_0\equiv \frac{1}{\tau_0} \simeq  170\left(\frac{1}{\eta}\right)~ {\rm kHz},  \nonumber
\ee
where we use $c_s\simeq 3$ km/s for speed of sound in rocks.   
 For large distances $r$ our  estimate for the frequency becomes 
\be
\label{frequency-underground}
\nu (\bar{x}^{\downarrow}) \sim\frac{\nu_0}{\sqrt[4]{\bar{x}^{\downarrow}}}
\sim    3.5 \left(\frac{1}{\eta}\right)^{\frac{3}{4}}\cdot  
\left( \frac{100\, {\rm km}}{r}\right)^{\frac{1}{4}} {\rm kHz},   
\ee
which is almost three  orders of magnitude higher than the frequency of the infrasound emitted by AQNs in atmosphere (\ref{frequencyAQN}).

In the estimate (\ref{SolidDeltaP})  above we assumed that the absorption can be neglected\footnote{In analogous estimate studied in previous Section \ref{atmosphere} for  the infrasound produced by AQN in the air  this  assumption is well- justified. One can easily convince yourself that the estimate  (\ref{AirDeltaPnumber})   is practically unaffected by absorption   on the distance well above 100 km.}. We now estimate the corresponding attenuation effects to support our assumption.  
To proceed  with estimates we note that  the sound absorption length scales as $\propto \nu^{-2}$.    A proper estimation of the absorption effects must include  the integration over distance where sound wave propagates  since the frequency depends on the distance according to (\ref{frequency-underground}) as $\nu \propto  r^{-1/4}$ and the absorption length  $l \propto \nu^{-2} \propto  r^{1/2}$. As a result, 
formula (\ref{SolidDeltaP})  will receive an additional exponential factor  which describes the suppression of the sound intensity  (intensity $\propto \delta p^2$)  due to the     absorption  of the sound wave: 
\be
\label{soundAbsorption}
 \exp(-X),\,\,~~~{\rm where}   ~~~X\approx 2 \frac{[L^{\downarrow} r]^{\frac{1}{2}}}{l_0\eta^2},
\ee
where $l_0$ is the absorption length for the initial frequency $\nu_0$.
For a numerical  estimate of the blast wave absorption  we may use detailed data on the sound absorption in sea water \cite{soundabsorption}.  The absorption length for sound with $\nu_0\simeq 170 $\,kHz is $l_0\simeq 0.1 $\,km, and we arrive at an estimate for  $X \simeq \eta^{-3/2}$ for $r=$ 100\,km.
Since $\eta$  is probably larger than 1,  this gives us an indication that a significant part of the blast wave may reach a detector on the distance up to 100 km.

The absorption of the blast wave is even smaller in water where the sound frequency is expected to be significantly smaller.  For the photon energy $\sim 100$\,keV the absorption length in water is 4.15\,cm, so we assume  $L^{\downarrow}=4.15 \eta $\,cm.  The speed of sound in water is $c_s$=1.5\,km/s, so we have  our  estimate for frequency in water  $\nu_0\simeq$  36 KHz and  
\be
\label{frequency-underwater}
\nu (\bar{x}^{\downarrow}) \sim\frac{\nu_0}{\sqrt[4]{\bar{x}^{\downarrow}}}
\sim     \left(\frac{1}{\eta}\right)^{\frac{3}{4}}\cdot  
\left( \frac{100\, {\rm km}}{r}\right)^{\frac{1}{4}} {\rm kHz}.   
\ee
The sound absorption length in water is  $l_0\simeq 0.45 $ km and $X \simeq 0.2 \eta^{-3/2}$ for $r=$ 100 km. Thus, the absorption of the blast wave in water is insignificant.

 We conclude this section with the following comment. In case of conventional meteoroids all numerical factors entering the scaling relations such as (\ref{delta_p}) and (\ref{tau}) have been fitted to match with numerous observations. Therefore, we introduce into our AQN estimates  empirical parameters $\xi^{\downarrow}$ and $\eta$ which are very hard to compute from the first principles, but could be  fixed by observations.
 Further studies are needed  to collect more statistics of mysterious events when sound signatures are recorded without any traces in the synchronized optical monitoring systems.


\exclude{------------------------------------

 The annihilation energy underground is 3  orders of magnitude higher than in the air due to 3  orders of magnitude higher density.  It is also very important that the absorption length $L$ in solids is 3 orders of magnitude smaller than in the air due to higher density. Therefore, three orders of magnitude larger  energy  is released in six orders of magnitude smaller volume ($V \sim L^2$ in a cylinder), leads to nine orders of magnitude larger increase of temperature and explosion.

 The nugget internal temperature in solids is about 5 times higher than in the air, $T_{\rm nugget}\sim 200$ KeV (see Appendix). For photons with $\omega$=200 KeV the absorption length in silicon is  $L\approx 2$ cm. This gives deposited energy about $10^4$J/cm$^3$ which leads to an instantaneous increase of temperature by $\sim$ 3000 K. The increase of temperature may be even higher in a close vicinity of the AQN track due to the contribution of the low frequency photons with a very small absorption length. This looks like explosion producing a blast wave.    
 
 Firstly this means that AQN left noticeable cm-wide tracks and bigger cracks around them in rocks and ice which may be search for now in the old rocks and Antarctic ice.
 Second, the blast wave from AQN may be detected at large distances.   
  }    
\exclude{  

An important question is absorption sound length which is proportional to λ2, λ is the wave length. If λ ∼ 10 cm , frequency ∼ 30 KHz, for 30 cm 10KHz. In ice the absorption length is calculated to be 9 3 km at all frequencies above 100 Hz. For NaCl (rock salt) with grain size 0.75 cm, scattering lengths are calculated to be 120 and 1.4 km at 10 and 30 kHz, and absorption lengths are calculated to be 3 104 and 3300 km at 10 and 30 kHz. Existing measurements are consistent with theory. For ice, absorption is the limiting factor; for salt, scattering is the limiting factor. Both media would be suitable for detection of acoustic waves from ultrahigh energy neutrino interactions. This also looks good for 10 km AQN distance.

    \be
   \label{tau_AQN_r}
  {\tau}^{\downarrow}(\bar{x}^{\downarrow})\sim   \bar{\tau}^{\downarrow}_0 \sqrt[4]{\bar{x}^{\downarrow}}  ~~~,~~~~
  \bar{x}^{\downarrow}\equiv \frac{r}{\sqrt{\xi^{\downarrow}}R_0^{\rm AQN{\downarrow}}}~,   
   \ee
 where the fundamental period for the underground sound $\bar{\tau}^{\downarrow}_0$ and corresponding  frequency are given by
 \be
\label{frequency_r}
    \bar{\tau}^{\downarrow}_0 \simeq \frac{R_0^{\rm AQN{\downarrow}}}{c^{\downarrow}_s}~~,~~~~  \bar{\nu}^{\downarrow} (\bar{x}^{\downarrow}) \equiv   \frac{1}{ {\tau}^{\downarrow}(\bar{x}^{\downarrow})}.
\ee
 }

\exclude{
 Parameter $R_0^{\rm AQN}$ from (\ref{R_AQN}) now assumes the  following numerical value 
  \be
\label{R_AQN_r}
  R_0^{\rm AQN\downarrow}\equiv \sqrt{ \frac{E_l^{\downarrow}}{p_0}  }\sim  10 ~ \left( \frac{B}{10^{25}}\right)^{\frac{1}{3}}\left({\frac{n_{\rm rock}}{10^{24} ~{\rm cm^{-3}}}}\right)^{\frac{1}{2}}  {\rm m},~~~~
\ee
 which of course 30 times larger than $R_0^{\rm AQN}$ in atmosphere as the rate of annihilation is much higher. 
 
 Similar to our analysis for AQN in atmosphere we introduce   an unknown parameter  $\xi^{\downarrow}$ to account for the complicated physics which describes the transfer of x ray energy into the sound wave energy:
   \be
   \label{blast_r}
   E^{\downarrow}_{\rm blast}\equiv \xi^{\downarrow} \cdot E^{\downarrow}_l, ~~~ \xi^{\downarrow}\lesssim (10^{-3}-10^{-4}),  
   \ee  
    where we again assume that $\xi^{\downarrow}$ represents very small portion of the total released energy (\ref{E_r}), while the dominant portion of the energy will be emitted in form of the $x$ rays.} 

$\bullet$ We conclude this section with few important  comments. Our prediction for the overpressure (\ref{AirDeltaPnumber})  for a typical AQN event with $B\simeq 10^{25}$ at the infrasound frequency (\ref{frequencyAQN}) suggests that  the existing instruments such as ELFO are not sufficiently sensitive to 
detect such small   signals on the level of $\delta p\simeq 0.03$\,Pa. Nevertheless, some strong and rare events still can be recorded with existing technology. In particular, it is tempting to identify a single mysterious event  recorded by ELFO \cite{ELFO}  as the AQN event with very large baryon charge $B\simeq 10^{27}$, see    \ref{observations} with detail arguments. Here we just highlight this reasoning. 

 The radius of the nugget goes as $R\sim B^{1/3} $ which effectively leads to an increase in the  number of annihilation events for larger nuggets, which eventually releases higher  energy output  per unit length as Eq.\,(\ref{E_l}) states. Therefore, the intensity of the event scales correspondingly. The recorded by ELFO a powerful  explosion with overpressure  on the level  $\delta p\simeq 0.3$\,Pa is consistent with  the AQN annihilation event with  large $B\simeq 10^{27}$. Such intense events    are relatively rare ones  according to (\ref{eq:f(B)}) and (\ref{eq:D Nflux 3})    as the frequency of appearance is proportional to $f(B)\sim B^{-\alpha}$ with $\alpha\simeq (2-2.5)$.  It might be a part of an explanation of why this area has observed a single event in 10 years rather than observing similar events much more often.

  It is obvious that we need much more statistics 
for systematic studies of relatively  small  but frequent typical events with $B\simeq 10^{25}$. In the next section we present a possible design of an instrument which could be sufficiently sensitive to infrasound and seismic   signals  to fulfill this goal.  If our proposal turns out to be successful, one  can routinely  record a large number of such events which manifest themselves 
in the form of the infrasonic and seismic signals, while the    optical synchronized  cameras may  not see any light from these events. Infrasonic signals must be always accompanied by sound and seismic waves as discussed above, and which can be routinely recorded by conventional seismic stations.  These events should demonstrate the daily and annual modulations as the source for these events  is  the  dark matter   galactic wind.

   \section{Detection strategy and possible   instruments}\label{detection-strategy}
   As we mentioned in previous section, the sensitivity of the instruments  similar to ELFO is not sufficient to record relatively small  events in the atmosphere which occur approximately  once a day in an area of $(100~ {\rm km})^2$ with $\delta p\simeq 0.01$\,Pa and frequency $\nu\simeq 5$ Hz. 
    In this section we suggest several possible designs of instruments   which we think are capable of fulfilling  this role as sufficiently sensitive devices  to  detect the dark matter signals from small and common  AQNs with $B\sim 10^{25}$. 
   
    We start the overview with a promising recent development, Distributed Acoustic Sensing (DAS), which is becoming a conventional tool for seismic and other applications, see, for example, \cite{Parker,Daley,Jousset,Ajo-Franklin} and references therein.  The basic idea of these activities  can be explained  as follows.  It has been known for quite sometime that distributed optical fiber sensors are  capable of measuring the signals  at thousands of points simultaneously using  an unmodified optical fiber as the sensing element. The recent  development is that the DAS is capable of measuring strain changes at all points along the optical fiber at {\it acoustic frequencies}, which is crucial for our studies of the acoustic waves emitted due to the AQN passage. 
    
    The main element of the DAS technology is that  a pulse of light is sent into optical fiber   and, through scattering in the glass, a small amount of the incident light is scattered back towards the sensing unit. The key point is that   the DAS is capable of determining from  this  scattered light, a component which indicates changes in the local axial strain along the fibre. It has been shown that this technology is capable of detecting signals at frequencies as low as 8\,mHz and as high as 49.5\,kHz with sensitivity at the level of $\delta p\sim 0.1\,{\rm mBar}\approx 10 ~{\rm Pa}$ \cite{Parker} which is more than sufficient for our purposes for $\delta p$   as estimated in (\ref{SolidDeltaP}) and typical  
     frequencies in kHz band as estimated in (\ref{frequency-underground}). Furthermore, it has been shown that 
     using an amplifier chain one can extend the range of DAS unit to 82 km, while maintaining high signal quality  \cite{Parker}. Such a long range on the scale of $100$ km matches well with what is needed for AQN-passage detection. Indeed, we anticipate approximately one event per day per an area of $(100~ {\rm km})^2$ according to (\ref{eq:D Nflux 3}) reviewed in Section \ref{AQN-flux}. An important point is that such studies can in principle  detect not only the intensity and the frequency of the sound wave, but also the direction of the source. We note that   networks of optical-fiber telecommunication cables cover a significant part of the Earth's surface. 
     
     We anticipate that the main problem with DAS  will be  separation of the AQN signal from the seismic noise and numerous spurious events. We discuss one of the  possibilities to separate the signal from a much larger noise below. The main point is that the AQN signal must show the annual (\ref{eq:annual}) and daily (\ref{eq:daily}) modulations  characteristic of the dark matter galactic wind, in contrast with much more numerous and much  more intense random events.  
\exclude{  
\textcolor{green}{Another venue to attack the same problem is to use a fiber-optics-connected network of atomic clocks. 
It was previously suggested to use, for example, the trans-European fiber-linked network of nine ultraprecise laboratory clocks [see for example, \cite{DereviankoPospelov,sensors,Wcislo,Safronova,Roberts,Roberts2,Roberts3}] as a tool to search for
low-mass scalar dark matter. }

\textcolor{green}{The phase of an optical wave propagating over a distance in a fiber  accumulates effects of external perturbation and is highly sensitive to these perturbations. The changes of the phases are detected by a network of synchronised atomic clocks and optical cavities. The original idea was to detect transient effects produced by clumps of low-mass dark matter and by passing topological defects. Here, we suggest that such networks may also be sensitive to the blast waves produced by AQN. }
  
\textcolor{green}{The blast wave pressure $\delta p$ produces a minor change of the optical fiber index of refraction and changes the optical length, $\delta L_{\rm opt}$. This leads to a change in the phase of order  $\phi=\delta L_{\rm opt}/\lambda$, where the wavelength $\lambda \sim 500$ nm = $0.5 \cdot 10^{-6}$ m.   The relative accuracy of the best atomic clocks has surpassed $10^{-18}$, but this requires a long observation time. As an estimate of the best current sensitivity we may take $ \delta L \sim 10^{-19}$\,m as in LIGO and VIRGO gravitational wave detectors. This very high sensitivity indicates that the main problem here is the separation of the AQN signal from the seismic noise.}
  }
  One of possibilities to separate the signal from a much larger noise was  suggested in \cite{Budker:2019zka}.
  
  The basic idea of \cite{Budker:2019zka}
  can be explained as follows. 
 The AQN flux  is given by (\ref{eq:D Nflux 3}) if averaged over very long period of time, longer than a year. However, due to the relative motion and orienation  of the Sun,  Earth and the galaxy, the AQN flux, as that of any other dark mater particle,  receives the time-dependent factor  $A_{\rm (a)}(t)$   representing the annual   modulation which is  defined as follows:
   \begin{equation}
\label{eq:annual}
 A_{\rm (a)}(t)\equiv[1+\kappa_{\rm (a)} \cos\Omega_a (t-t_0)],
\end{equation}
 where $\Omega_a=2\pi\,\rm yr^{-1}\approx 2\pi\cdot\,32\,$nHz is the angular frequency of the annual modulation
 and   label $``a"$ in $\Omega_a $ stands for annual.   The $\Omega_a t_0$ is the phase shift corresponding to the maximum on June 1 and minimum on December 1 for the standard galactic DM distribution, see \cite{Freese:1987wu,Freese:2012xd}. 
 Similar daily modulations are also known to occur \cite{Liang:2019lya} and can be represented as follows:
 \begin{equation}
\label{eq:daily}
A_{\rm (d)}(t)\equiv[1+\kappa_{\rm (d)} \cos(\Omega_d t-\phi_0)],
\end{equation}
 where   $\Omega_d=2\pi\,\rm day^{-1}\approx 2\pi\cdot 11.6\,\mu$Hz is the angular frequency of the daily modulation, while $\phi_0$ is the phase shift similar to $\Omega_at_0$ in (\ref{eq:annual}). It   can be assumed to be constant on the scale of days. However, it actually slowly changes  with time due to the  variation of the direction of  DM wind   with respect  to the Earth. The modulation coefficients $\kappa_{\rm (a)}$ and $\kappa_{\rm (d)}$
 have been computed in the AQN model in \cite{Liang:2019lya}, see also  application of these results to analysis  of the seasonal variation observed by XMM-Newton  collaboration in X-ray bands \cite{Ge:2020cho}. 
 
 The idea advocated in \cite{Budker:2019zka} is to fit the data to the  modulation 
 formulae (\ref{eq:annual}) and (\ref{eq:daily}) even if the noise is large and exceeds the expected signal.  The key point here is the statistics factor and accumulation of the signal for a long period of time assuming that the noise can be treated as being   random  in contrast  with signal  being characterized by well defined frequencies $\Omega_a$ and $\Omega_d$.
    A hope is to  discover  the annual (\ref{eq:annual}) and    daily (\ref{eq:daily}) modulations by recording a large number of AQN events which represent the  dark matter  
   galactic wind in this specific model.

A specific signal from AQN tracks is very different from the ordinary seismic noise and Earthquakes. Therefore, AQN signals may, in principle,   be detected by an existing network of seismic stations.  In addition to optical-fiber based methods, it may be possible to search for the AQN-passage signatures also in the large volumes of existing historical data from networks of seismometers \cite{Iris}.

 We would like to briefly mention other possibilities for the AQN detection, see also relevant references in Section \ref{AQN-flux} leading to the  constraints (\ref{window}). 
  The AQN produce only a small amount of visible light as we already mentioned.  However, the emitted x-rays will be absorbed and heat the atmosphere along the track on scales of order $L$ as Eq.\,(\ref{lambda_AQN}) suggests.  It may produce vapor tracks along the AQN path.
  Therefore,  one may try to observe infrared radiation from AQN tracks using infrared telescopes being synchronized with infrasonic detectors and all-sky  cameras. Similarly, the
 AQN tracks also produce microwave and radio wave radiation which may be detected by radio telescopes which can be also synchronized with infrasonic and all sky  cameras.
 
  Furthermore, while an AQN itself  is  only $R\simeq 10^{-5}$\,cm in size, nevertheless, it may leave  larger and  noticeable    cracks   along its path in solids as instantaneous defect creation and temperature increase 
 occur on a cm scale according to (\ref{temperature}). The cracks could be sufficiently large to be observed. A search for AQN annihilation tracks could also be performed in old rocks and Antarctic ice.

Finally, one more possibility to study the long ranged signals which could be   produced by the AQN traversing the Earth atmosphere  is to search for   signals   similar to 
the ones which are normally attributed  to  the ultra  high energy cosmic rays. One should consider specifically  the area   under the thunderstorm conditions\footnote{such condition occur on average for $1\%$ of the time. The  estimation   is based on compilation of the annual thunderstorm duration from 450 air weather system in USA, see \cite{Zhitnitsky:2020shd} for references.}   which could generate enormous enhancement factor in the intensity of the signal as argued in \cite{Zhitnitsky:2020shd}. 
 In fact,  the recent unusual events recorded by  Telescope Array Experiment might be precisely those types of events, see     \cite{Zhitnitsky:2020shd} for the details.

   \section{Conclusion  \label{conclusion}} 
   The main results of the present work can be summarized as follows:\\
   1.  We argue that an AQN propagating in the Earth atmosphere generates infrasonic waves. We estimated the intensity (\ref{AirDeltaPnumber}) and spectral features (\ref{frequencyAQN}) of these waves
    for a typical AQN event with $B\simeq 10^{25}$;\\  
    2. We also performed similar estimates for an AQN propagating inside the Earth;\\
    3. We propose a detection strategy to search 
for a signal generated by  a typical relatively small AQN event with $B\simeq 10^{25}$ by using Distributed Acoustic Sensing as existing instruments are not sufficiently sensitive to detect such signals;\\  
4. Specific signals from AQN passage may also be detected with a variety of alternative techniques, for instance, with an existing network of seismic stations (or even by analyzing the already existing data) or by analyzing specific correlations under the thunderstorm conditions which may mimic the Ultrahigh Energy Cosmic Ray events, as mentioned  at the end of  Sec.\,\ref{detection-strategy};\\
   5.We further speculate  that 
     the mysterious explosion which occurred on July 31st 2008 and which was properly recorded   by the dedicated   Elginfield Infrasound Array     \cite{ELFO}  
   might be a good candidate for an AQN-annihilation event with very large $B\simeq 10^{27}$ as our basic estimates for the overpressure 
(\ref{numerics-1}) and the frequency (\ref{numerics-2})  are amazingly close to the  signal recorded by ELFO.  
     
       One should emphasize that our estimates are based on the parameters of the AQN model which were fixed long ago for completely different reasons, namely, to explain various different phenomena occurring in drastically different environments, as reviewed in Sec.\,\ref{AQN-flux}.

 Why should one take  this (AQN)  model seriously? 
 A simple answer is as follows.   Originally, this model  was invented to  explain 
 the observed relation $\Omega_{\rm DM}\sim \Omega_{\rm visible}$  where  the   ``baryogenesis" framework   is replaced with a ``charge-separation" paradigm,  as reviewed  in the Introduction and in Sec.\,\ref{AQN-flux}. 
 This model is shown to be   consistent with all available cosmological, astrophysical, satellite and ground-based constraints, where AQNs could leave a detectable electromagnetic signature as discussed in the Introduction and Sec.\,\ref{AQN-flux}, with one and the  {\it same set }  of parameters. 
  Furthermore, it  has been also shown   that the AQNs   could be formed and  could  survive the unfriendly environment of the  early Universe. Therefore, the AQNs deserve to be considered a viable DM candidate.   Finally, the same AQN framework    may also explain a number of other (naively unrelated) observed phenomena as mentioned in Sec.\,\ref{AQN-flux}.  
   
 \exclude{
  Our identification (\ref{identification}) between mysterious skyquakes and the AQN annihilation events, if confirmed by future studies,    would be the    {\it first~direct }  (non-gravitation) evidence which reveals  the nature of the DM, in  contrast with {\it indirect} observations mentioned above. 
} 
  
\section*{Acknowledgments}
  The authors are grateful to Mark\,G.\,Raizen, Vincent Dumont, Nataniel Figueroa Leigh  and Jonathan\,B.\,Ajo-Franklin  for the discussion of fiber-networks for acoustic detection. The work of DB was supported in part by the DFG Project ID 390831469:  EXC 2118 (PRISMA+ Cluster of Excellence). DB also received support from the European Research Council (ERC) under the European Union Horizon 2020 Research and Innovation Program (grant agreement No. 695405), from the DFG Reinhart Koselleck Project and the Heising-Simons Foundation.
 The work of  VF is supported by the Australian Research Council and the Gutenberg Fellowship.
  The work of AZ   was supported in part by the National Science and Engineering Research Council of Canada.

 \appendix
  \section{Theory  confronts observations}\label{observations} 
   
  As we mentioned in the main text of the paper   our prediction for the overpressure (\ref{AirDeltaPnumber})  for a typical AQN event with $B\simeq 10^{25}$ at the infrasound frequency (\ref{frequencyAQN}) suggests that  the predicted signal is too weak to detect by existing instruments such as ELFO. We also mentioned that  some strong and intense   events may still occur and can be recorded with existing technology. 
  However, the frequency of appearance for such intense events should be very low. 
  
  In particular, it is tempting to identify a single mysterious event  recorded by ELFO \cite{ELFO}  as the AQN event with very large baryon charge $B\simeq 10^{27}$. 
  
\begin{figure}
    \centering
    \includegraphics[width=0.7\linewidth]{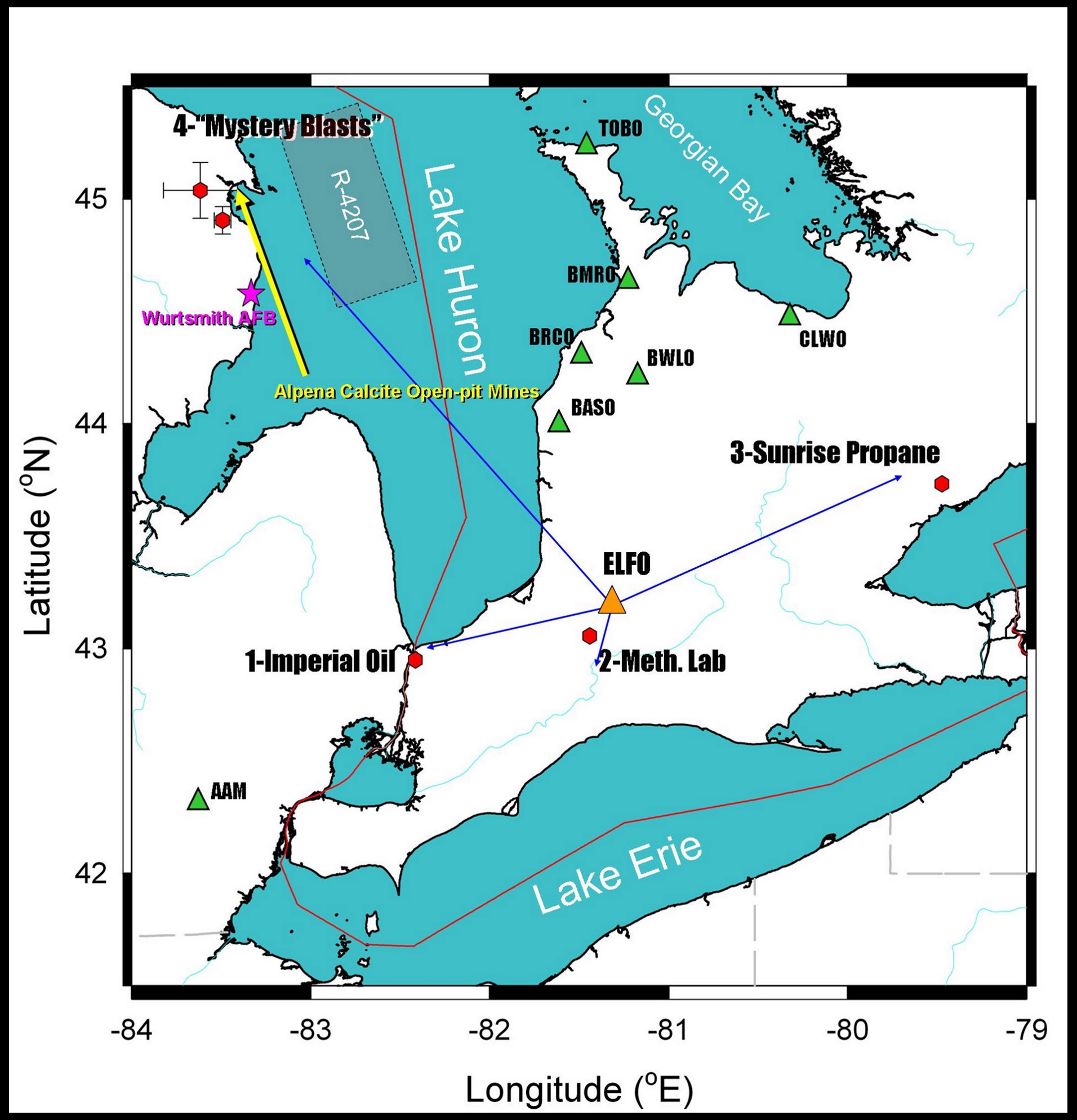}
    \caption{Location of ELFO and seismic stations in the area, adopted from \cite{ELFO}. One degree along the latitude   corresponds to 112 km. i.e. $1^0\approx  112$ km, while  along the longitude $1^0\approx 82$ km.  It explains our benchmark 300 km in eqs. (\ref{numerics-1}) and (\ref{numerics-2}) which covers the relevant area shown on the map. The green triangles represent the seismic  stations in the area. Red symbols with error bars represent the position of the blasts (with errors) assuming quarry-type explosions, other red symbols with blue lines directed to them show the  directions from ELFO to these potential sources of the explosions (they  have been considered but ruled out as the sources).}
    \label{mysterious-1}
\end{figure}

  The goal of this Appendix is to present the arguments
 suggesting that the mysterious explosion which occurred on July 31st 2008 and which was properly recorded   by dedicated   Elginfield Infrasound Array    \cite{ELFO}   
   is consistent with the AQN annihilation events. 
   Localization of the source position,    Elginfield Infrasound Array  (ELFO) and seismic stations  are  shown on Fig.\ref{mysterious-1} adopted from 
    \cite{ELFO}.
    
\begin{figure}
    \centering
    \includegraphics[width=0.7\linewidth]{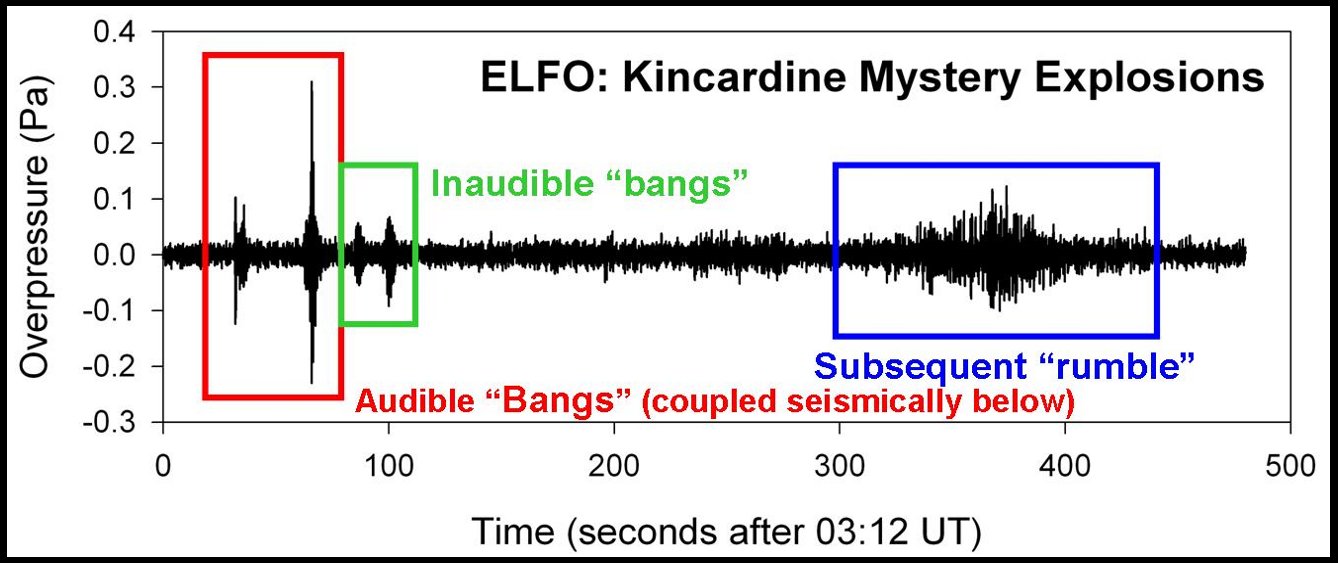}
    \caption{Infrasound impulses as recorded by ELFO,  adopted from \cite{ELFO}.}
    \label{mysterious-3}
\end{figure}

\begin{figure}
    \centering
    \includegraphics[width=0.7\linewidth]{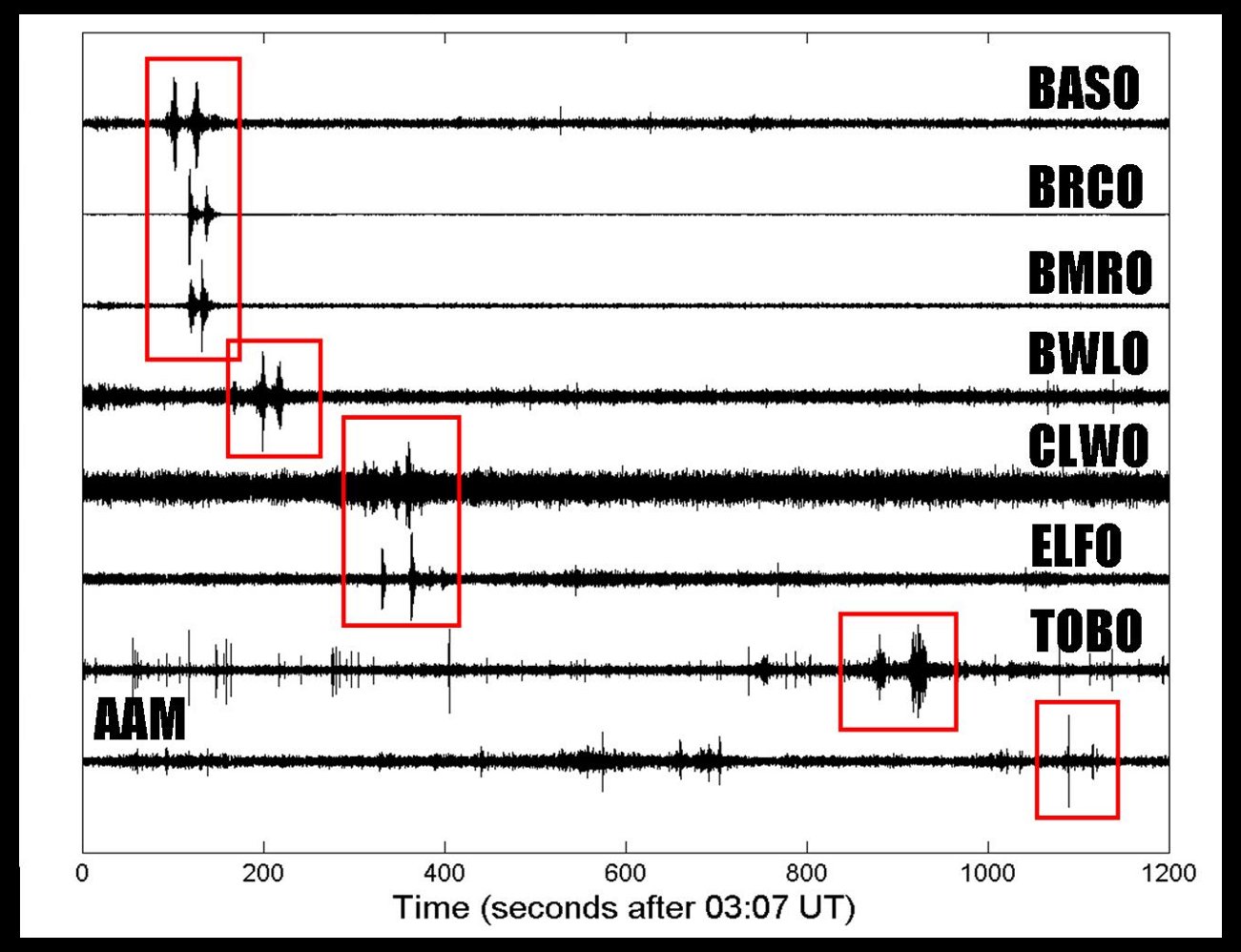}
    \caption{Impulses as observed by seismic  stations  in the area, adopted from \cite{ELFO}.}
    \label{mysterious-2}
\end{figure}

   The sounds, as reported by   residents of Kincardine, Ontario, Canada were apparently loud enough to rattle windows and objects on walls.
   An important  point here is that the infrasound detection  associated with this sound shock was recorded by ELFO as  presented in Fig.\,\ref{mysterious-3} with a typical overpressure 
   $\delta p\sim 0.3~{\rm Pa}$. These observations (along with non-observations in the all-sky camera network) ruled out a meteor source, as well as operations at the Bruce Nuclear Power Plant, while Goderich salt mine logs eliminated it as a source  \cite{ELFO}.   Furthermore, a  local airport radar reported no aircraft in the area at the time. The impulses were also observed seismically as ground coupled acoustic waves around South Western Ontario and Northern Michigan as shown on Fig. \ref{mysterious-2}.  One should emphasize  that the seismic stations which record ground coupled acoustic waves may detect signals before ELFO because the speed of sound in a solid is much higher than in the atmosphere. Furthermore, the propagation  of seismic waves is complicated and depends on  the  geological structure, which is specific to the local area.  There are  also different propagation speeds for  transverse and longitudinal waves.
   \exclude{It has been claimed  that} The analysis  of the seismic stations \cite{ELFO} suggests that two blasts (if interpreted as quarry-type explosions) are localized  in the area denoted with the red symbols with error bars on Fig.\ref{mysterious-1}.
   It should be contrasted with the infrasound signal detected by ELFO which arrives from the atmosphere from a different location and propagates with a much lower speed. 
   
  Now we are in a position to apply the scaling behaviour (\ref{AirDeltaPnumber}) to see if  the mysterious event recorded by ELFO 
  might have resulted from the AQN annihilation events along its trajectory when it crosses the atmosphere in  this area. Our estimate  (\ref{AirDeltaPnumber}) suggests that the overpressure $ {\delta p}$ assumes the following numerical value
  at $r\simeq 300 $ km:
  \be
  \label{numerics-1}
 \delta p\approx 0.3\,{\rm Pa} \left( \frac{B}{10^{27}}\right)^{2/3}\left( \frac{300\, {\rm km}}{r}\right)^{3/4},~~  ~~~
 \ee
 where we  choose the benchmark for $r=300$ km corresponding to a typical distances in the area shown on Fig. \ref{mysterious-1}, and $B=10^{27}$ to bring the numerical coefficient close to the   measured  value  ${\delta p}\simeq 0.3$ Pa as recorded by ELFO,  see Fig. \ref{mysterious-3}. Estimation for the intensity (\ref{numerics-1}) 
    is consistent with observation ${\delta p}\simeq 0.3$ Pa if one assumes that the mysterious explosion  had resulted from the AQN annihilation event of a relatively large size with the baryon charge $B\simeq 10^{27}$.  We shall support this interpretation at the end of this Appendix by analysing the frequency of appearance  of such large sized nuggets.

    Another parameter which characterizes the acoustic shock is the frequency determined by the scaling formula (\ref{frequencyAQN}).
   Assuming the same numerical parameters as before we arrive at the following numerical estimate: 
    \be
\label{numerics-2}
\nu (\bar{x}) 
\sim  \frac{\nu_0}{{\bar{x}^{1/4}}}\sim 5\, {\rm Hz} \left( \frac{300\, {\rm km}}{r}\right)^{1/4}, 
\ee
which is precisely in the range of the highest sensitivity  of ELFO where the noise levels are: $10^{-4} {\rm {Pa^2}/{ Hz}}$ for the 10 Hz frequency band and 
$10^{-3} {\rm  {Pa^2}/{ Hz}}$ in  the 1 Hz frequency band  \cite{ELFO}.

Now our task is to estimate the relevant frequency for the sound emitted in the underground  rocks.   The corresponding expression for the frequency at $r\simeq 300$ km is determined by the scaling relation (\ref{frequency-underground}) 
and it is given by:
 \be
\label{numerics-3}
\nu (\bar{x}^{\downarrow}) \sim\frac{\nu_0}{\sqrt[4]{\bar{x}^{\downarrow}}}
\sim    2.5 \left(\frac{1}{\eta}\right)^{\frac{3}{4}}\cdot  
\left( \frac{300\, {\rm km}}{r}\right)^{\frac{1}{4}} {\rm kHz}.   
\ee
 The estimate for the underground frequency of Eq.\,(\ref{numerics-3}) should be contrasted with the estimate of Eq.\,(\ref{numerics-2}) for the atmosphere. The basic observation is that the acoustic waves in the atmosphere are in the infrasonic frequency range, while underground, they are in the sound frequency band, as already mentioned. 
    
Several comments are in order. First of all, our proposal demonstrates  a qualitative consistency with   the mysterious event recorded by ELFO on July 31st 2008. Indeed,  as we mentioned above the AQNs do not emit directly the visible light, see 
\ref{sect:spectrum}. It should be  contrasted with conventional meteors which directly emit the visible light being consistent with  black  body radiation  spectrum.  This observation implies that the AQNs cannot be observed by conventional optical monitoring.   
The mysterious event recorded by ELFO on July 31st  falls into this category   because it was observed by  infrasound instruments  and  not observed  by the   optical   synchronized cameras. 

Another qualitative  comment goes as follows.     The frequency (\ref{numerics-3})  of the sound 
from  the underground blast falls in human  hearing  range.   Therefore, it is also consistent with the fact that residents of nearby Kincardine could hear the sound (rather than infrasound) signal which originates from the underground with frequency (\ref{numerics-3}).  It is also consistent with  seismic observations which are  correlated\footnote{The corresponding  bangs are classified as seismically coupled ``Audible Bangs".} with infrasound impulses recorded by ELFO. As discussed above, the atmospheric and underground acoustic waves emitted by the same AQN are always accompanied by each other as they originate from  one and the same AQN propagating from outer space through the atmosphere and the Earth.  

How we should interpret two blasts recorded in the acoustic and infrasonic frequency bands?  
The answer depends on the theory of formation of the acoustic  waves as a result of ultrasonic motion with very large Mach number which of course is not developed yet.
The complicated structure of the signal may, in principle, be explained by the fact that there are actually four waves moving with different speed and originating from different points,  which arrive at a different time: atmospheric wave (slow), longitudinal and transverse waves coming from underground,  and surface wave. In addition, there are reflected waves (echo). 
The answering all these hard questions obviously requires an analysis of large number of   events, and  obviously cannot be  accomplished at the present time
with a single recorded event. 
However,   the basic characteristics  such as the frequency and over-pressure for infrasonic signal are consistent with ELFO record.

Furthermore,   our estimations are also consistent on the quantitative level  with  the mysterious event recorded by ELFO on July 31st 2008. Indeed,  our estimates for overpressure (\ref{numerics-1}) and the frequency estimate (\ref{numerics-2})  are consistent with description   \cite{ELFO} represented on  Fig. \ref{mysterious-3}. 
This obviously should be considered as a strong support of our identification   between a rare intense AQN event and the mysterious explosion  recorded by ELFO.

\exclude{
We also comment on a different event that happened in Alabama sky in 2017 when the sound was heard across 15 counties. There were no related meteor-activity reports. There were, however, reports of {\it vapor} trails \cite{NASA}. Our original comment here is that the presence of the {\it vapor}  is expected and in fact predicted by the proposal (\ref{identification}) because most of the released energy (\ref{E_l}) will heat the surrounding material, while a small portion of the energy will be released in the form of the sound (infrasound) waves as explained in the text.  There will be no significant emission in visible bands. Therefore, it is not a surprise  that  there were no reports on meteor activity in that event but there   were reports of {\it vapor} trails \cite{NASA}.
 }
 
  Our final comment is related to the energetics and frequency of appearance of such mysterious events interpreted in terms of the AQNs. As we already mentioned at the end of Section  \ref{estimates} we interpret  a relatively high overpressure (\ref{numerics-1}) at the level of $\delta p\simeq 0.3$\,Pa as a result of hitting  of a  sufficiently large AQN with $B\simeq 10^{27}$ which represents our explanation of why this area has observed a single event in 10 years rather than observing similar events much more often. 
  \exclude{
  The radius of the nuggets goes as $R\sim B^{1/3} $ which effectively leads to an increase in the  number of annihilation events for larger nuggets, which eventually releases higher  energy output  per unit length as Eq.\,(\ref{E_l}) states. Such events with large $B\simeq 10^{27}$ are relatively rare according to (\ref{eq:D Nflux 3}) and (\ref{eq:f(B)}) as the frequency of appearance is proportional to $f(B)\sim B^{-\alpha}$ with $\alpha\simeq (2-2.5)$.

 We argued that a single mysterious event properly recorded by ELFO on July 31st 2008 supports our proposal on identification (\ref{identification}) between dark matter AQN events and skyquakes. We need much more statistics 
to convincingly support this identification. In the next section we present a possible design of an instrument which could be sufficiently sensitive to infrasound signals such that much smaller (than mysterious event recorded by ELFO on July 31st 2008)  but much more frequent  events with $B\simeq 10^{25}$ could be recorded.  In this case, we can systematically record a large number of such events which manifest themselves 
in the form of the infrasonic signals, while the    optical synchronized  cameras may  not see any light from these events. Infrasonic signals must be always accompanied by sound waves as discussed above, and which can be routinely recorded by conventional seismic stations, similar to the ones presented on Fig.\,\ref{mysterious-2}.  These events should demonstrate the daily and annual modulations as the source for these events  is  the  dark matter   galactic wind. 
}

 \section{AQN emission spectrum}\label{sect:spectrum} 
 The goal of this Appendix is to overview the spectral characteristics of the AQNs as a result of annihilation events 
 when the nugget enters the Earth atmosphere. The corresponding computations have been carried out in \cite{Forbes:2008uf}
 in application to the galactic environment with a typical density of surrounding visible baryons of order $n_{\rm galaxy}\sim 300 ~{\rm cm^{-3}}$ in the galactic center. We review  these computations with few additional elements which must be implemented for Earth's atmosphere when 
 typical density of surrounding baryons is much higher $n_{\rm air}\sim 10^{21} ~{\rm cm^{-3}}$.
 
The spectrum of nuggets at low 
temperatures was analyzed in \cite{Forbes:2008uf} and was found to be,
\begin{multline}
  \label{eq:P}
  \frac{\d{F}}{\d{\omega}}(\omega) = 
  \frac{\d{E}}{\d{t}\;\d{A}\;\d{\omega}}
  \simeq
  \frac{1}{2}\int^{\infty}_{0}\!\!\!\!\!\d{z}\;
  \frac{\d{Q}}{\d{\omega}}(\omega, z)
  \sim \\
  \sim
  \frac{4}{45}
  \frac{T^3\alpha^{5/2}}{\pi}\sqrt[4]{\frac{T}{m}}
  \left(1+\frac{\omega}{T}\right)e^{-\omega/T}h\left(\frac{\omega}{T}\right),
\end{multline}
where $Q(\omega, z)\sim n^2(z, T)$ describes the emissivity per unit volume from the electrosphere   characterized by the density  $n(z, T)$, where $z$ measures the distance from the quark core of the nugget. In Eq.\,(\ref{eq:P}) a complicated function $h(x)$ can be well approximated as 
\begin{equation}
  h(x) = \begin{cases}
    17-12\ln(x/2) & x<1,\\
    17+12\ln(2) & x\geq1.
  \end{cases}
\end{equation}
Integrating over $\omega$ contributes a factor of
$T\int\d{x}\;(1+x)\exp(-x)h(x)\approx 60\,T$, giving the total surface
emissivity:
\begin{equation}
  \label{eq:P_t}
  F_{\text{tot}} = 
  \frac{\d{E}}{\d{t}\;\d{A}} = 
  \int^{\infty}_0\!\!\!\!\!\d{\omega}\;
  \frac{\d{F}}{\d{\omega}}(\omega) 
  \sim
  \frac{16}{3}
  \frac{T^4\alpha^{5/2}}{\pi}\sqrt[4]{\frac{T}{m}}.\\
\end{equation}
Although a discussion of black-body radiation is inappropriate for
these nuggets (for one thing, they are too small to establish thermal
equilibrium with low-energy photons), it is still instructive to
compare the form of this surface emissivity with that of black-body
radiation $F_{BB}=\sigma T^4$:
\begin{equation}
  \label{eq:BB}
  \frac{F_{\text{tot}}}{F_{BB}} \simeq
  \frac{320}{\pi^3}\alpha^{5/2}\sqrt[4]{\frac{T}{m}}.
\end{equation}
At $T=1\eV$ which was an appropriate internal nugget's temperature for  the galactic environment, the emissivity $F_{\text{tot}} \sim 10^{-6} F_{BB}$ is much smaller
than that for black-body radiation.  As we discuss below a typical  internal nugget's temperature when AQN enters the atmosphere is of order of  $T=10~ {\rm keV}$ which results in the emissivity $F_{\text{tot}} \sim 10^{-5} F_{BB}$ for such high temperatures. 

 A typical internal temperature of  the nuggets can be estimated from the condition 
 the radiative output of equation (\ref{eq:P_t}) must balanced the flux of energy onto the 
nugget  due to the annihilation events. In this case we may write, 
\begin{equation}
\label{eq:rad_balance}
(4\pi R^2) \frac{16}{3}
  \frac{T^4\alpha^{5/2}}{\pi}\sqrt[4]{\frac{T}{m}}
\simeq \kappa (\pi R^2) 2~ {\rm GeV}n_{\rm air} v_{\rm AQN},
\end{equation} 
where the left hand side accounts for the total energy radiation from the nuggets' surface per unit time,
while the right hand side  accounts for the rate of annihilation events when each successful annihilation event of a single baryon charge produces $\sim 2m_pc^2\simeq 2~{\rm GeV}$ energy. In Eq.\,(\ref{eq:rad_balance}) we assume that  the nugget is characterized by the geometrical cross section $\pi R^2$ when it propagates 
in environment with local density $n_{\rm air}$ with velocity $v_{\rm AQN}\sim 10^{-3}c$.

The factor $\kappa$ is introduced to account for the fact that not all matter striking the nugget will 
annihilate and not all of the energy released by an annihilation will be thermalized in the nuggets (e.g. some portion of the energy will be released in form of the axions and neutrinos).
As such $\kappa$ encodes a large number of complex processes including the probability that 
 not all atoms and molecules   are capable to  penetrate into the 
color superconducting phase of the nugget to get annihilated. 
 In a neutral environment when no 
long range interactions exist the value of $\kappa$ cannot exceed $\kappa \sim 1$ which would 
correspond to the total annihilation of all impacting matter into to thermal photons. The high probability 
of reflection at the sharp quark matter surface lowers the value of $\kappa$. The propagation of an ionized (negatively charged) nugget in a  highly ionized plasma    will 
 increases the effective cross section, and therefore value of  $\kappa$ as discussed in \cite{Raza:2018gpb} in application to the solar corona heating problem. 
 
 For the neutral environment (such as Earth's atmosphere) and relatively low temperature when the most positrons from electrosphere remain in the system  the parameter $\kappa$ should assume values close to unity, i.e. $0.1\lesssim\kappa\lesssim 1$. In this case, 
 assuming that $0.1\lesssim\kappa\lesssim 1$ one can estimate a typical internal nugget's temperature in the Earth atmosphere:
 \be
 \label{T}
 T\sim 40 ~{\rm keV} \cdot \left(\frac{n_{\rm air}}{10^{21} ~{\rm cm^{-3}}}\right)^{\frac{4}{17}} \kappa^{\frac{4}{17}}.
 \ee 
Thus, in the air $T\simeq$ (20-40) KeV, depending on parameter $\kappa$.
In case of  solids the temperature must be higher because of higher density. This leads to a stronger AQN-ionization. This attracts more positively charged ions from surrounding material, which consequently may increase the rate of annihilation (effectively increasing $\kappa$).  All these effects are very complicated at large $T$ and deserve a separate study. In the present work for the order of magnitude estimates we adopt the previous  value of $0.1\lesssim\kappa\lesssim 1$ for the  case of  solids  as well, which would correspond to   $T \simeq$ (100-200) KeV. All the uncertainties  related to $\kappa$ do not modify our qualitative discussions in this work. There is much more important  element of the spectrum which drastically affects  the observational consequences of the emission. It is   a relatively  large magnitude of plasma frequency $\omega_p$ of the electrosphere at $T\neq 0$ when the  temperature reaches  $T\simeq$ (20-40) KeV in atmosphere as estimated above. 

Indeed, there are few additional elements which should be taken into account for Earth's atmosphere in comparison with  original computations \cite{Forbes:2008uf,Forbes:2009wg} applied to very 
dilute galactic environment with much lower temperatures $T\simeq 1 $ eV. These effects do not modify the basic scale (\ref{T}).
However, these additional elements strongly affect the spectrum at the lower  frequency bands.  In particular   the visible portion of the spectrum at $\omega\sim 1 $eV demonstrates a dramatic suppression. It  has some profound consequences for the present work as discussed in the main  body of the text. In particular, it implies that the AQNs do not emit the visible light with  $\omega\sim 1 $eV, in huge contrast with conventional meteors and meteorites which are normally characterized by  strong emission in the visible frequency bands through sputtering and ablation \cite{Silber-review, Silber-optical-1, Silber-optical-2}. 

We start our analysis on additional elements to be implemented with 
 the plasma frequency $\omega_{p}$ which characterizes the propagation of photons in the ionized plasma, which represents 
 the electrosphere for our AQN system.  The  $\omega_{p}$ can be thought as an effective mass for photon: only photons with the energy larger than this mass can propagate outside of the system, while  photons with $\omega<\omega_{p}$ can only propagate for a short time/distance $\sim\omega_{p}^{-1}$ before they get absorbed by the plasma. For our estimates we shall use a conventional non-relativistic expression  for $\omega_{p}$:
\begin{equation}
\label{eq:omega}
    \omega_p^2(z, T)=\frac{4\pi\alpha n(z, T)}{m},  
\end{equation}
where   the positron density $n(z, T)$ in electrosphere   in
the nonrelativistic mean-field approximation  has been computed in \cite{Forbes:2008uf,Forbes:2009wg}:
\begin{equation}
  \label{eq:n_r}
  n(z,T)\simeq \frac{T}{2\pi\alpha}\cdot \frac{1}{(z+\bar{z})^2}, ~~ \frac{1}{\bar{z}}
    \simeq \sqrt{2\pi\alpha }  m   \sqrt[4]{\frac{T}{m}},
\end{equation}
where $\bar{z}$ is a constant of integration being  determined by  
appropriate boundary condition deep inside the nugget's core. 
 Important  implication of the plasma frequency $\omega_p (z, T)$ 
is that the very dense regions in electrosphere essentially do not emit the photons with $\omega\lesssim  \omega_p (z, T)$.

 There is another effect which further suppresses  the emission of low energy photons. 
 It is related to the ionization processes when the AQN assumes a sufficiently large negative charge due to the $T\neq 0$ 
 as estimated above (\ref{T}). Essentially it affects all  loosely bound positrons which  will be completely stripped off by high temperature, while more strongly bound positrons will be less affected by the same temperature. 
 The corresponding effect  leads to very strong suppression of low energy photons with $\omega \ll T$ as  loosely bound positrons 
 represent the main source  of low frequency photons.  
 
\begin{figure}
    \centering
    \includegraphics[width=1\linewidth]{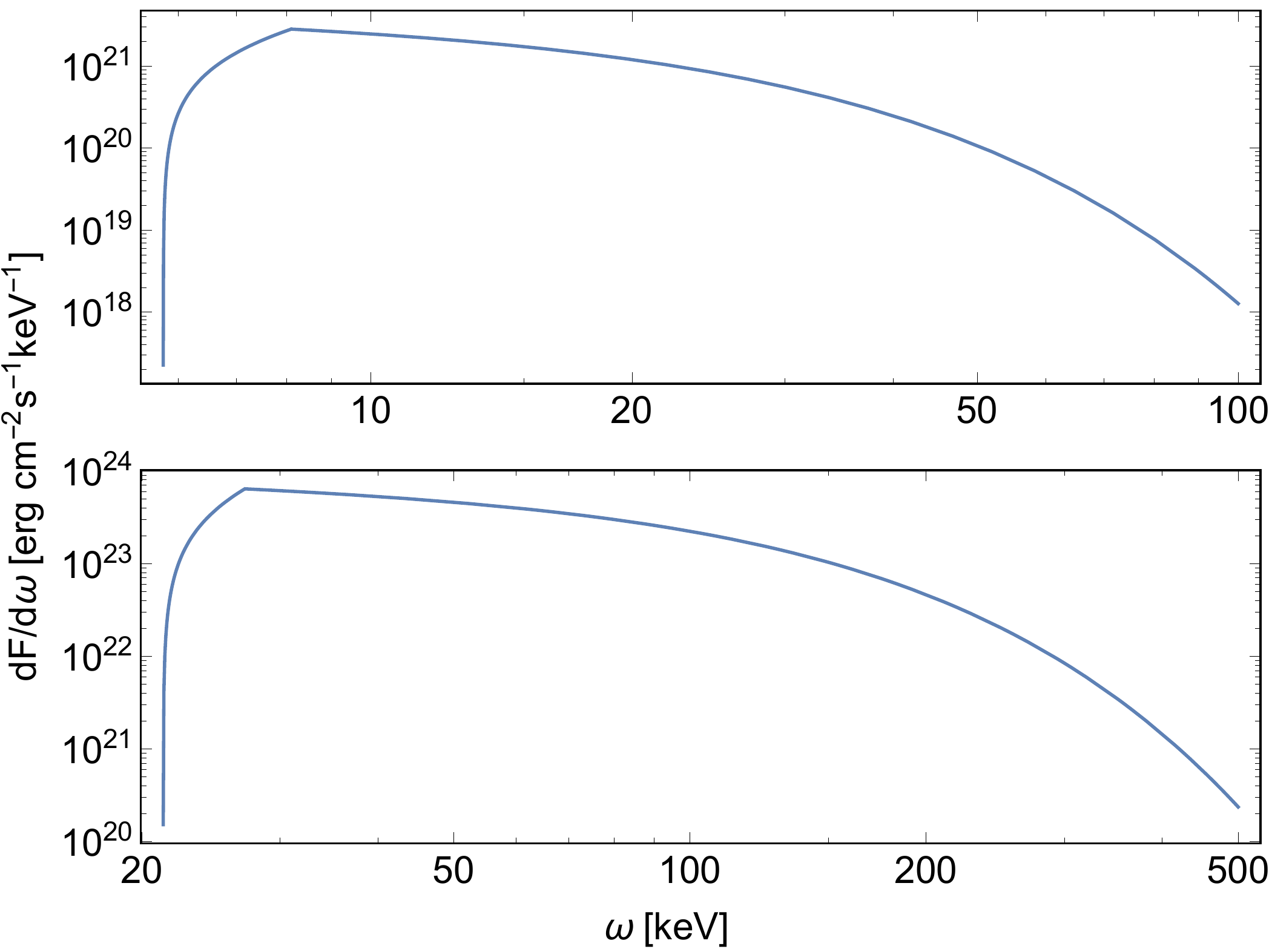}
    \caption{The spectral surface emissivity of a nugget with   the suppression effects at $\omega\ll T$ as discussed in this Appendix.      The top plot  corresponds to  $T=10$ keV  while the bottom plot corresponds to $T=50 $ keV.}
    \label{spectrum}
\end{figure}

 Both these effects have been implemented in Eq.\,(\ref{eq:P}) by performing numerical computation of the integral $\int \d z$ over electrosphere region with corresponding modifications of the positron density $n(z, T)$ and inserting  $\omega_p(z, T)$ as discussed above.   We present the corresponding results for these numerical studies  on Fig.\ref{spectrum} for $T=10$ keV and $T=50$ keV.    These values for the temperature essentially cover the  most relevant window (\ref{T}) for the present analysis   which dealt with   AQN emission in atmosphere. 
 
 \exclude{
To accommodate this feature in our simplified analysis we describe $k$ as the step function: 
\begin{equation}
\label{k}
k(z,T)=
\begin{cases}
0 ~~\text{if }~~ z\geq z_1~   \\
k(T)~~ \text{if} ~~ z < z_1
 \end{cases},
\end{equation}
where ``loosely bound" positrons are defined as positrons with $p^2\lesssim 2m T$, which mostly resided in the region of electrosphere with 
   $z_{1}(T)\simeq \hbar/p\geq \frac{1}{\sqrt{2m_{e}T}}$ is defined  
\begin{equation}\label{eq:z_1}
    z_{1}(T)\simeq\frac{1}{\sqrt{2m_{e}T}}.
\end{equation}
 }

   Few comments are in order. First of all, as one can see from Fig.\ref{spectrum}   the spectrum 
   is almost flat in the region $\omega\lesssim T$ which is the  direct manifestation of the well known soft photon theorem 
   when the emission of the photon with frequency $\omega$ is proportional to $\d\omega/\omega$.      
  For large $\omega\gg T$ the exponential suppression $\exp(-\omega/T)$ becomes the most important element of the spectrum. The     emission is strongly suppressed 
   at very small $\omega\simeq \omega_p\ll T$.    The strong suppression of  the spectrum with $\omega\ll T$  has profound phenomenologically   consequences:  drastic intensity drop  at small $\omega\ll T$ implies that  that the luminosity  of the visible light from AQN  with $\omega\sim (1-10)$ eV is  strongly suppressed in comparison with $X$-ray emission. This strong suppression is entirely due to the two effects mentioned above: the  presence of the plasma frequency in electrosphere (\ref{eq:omega}) and complete stripped off of the loosely bound positrons.   It implies that the AQNs cannot be observed by conventional optical monitoring  as AQNs are not   accompanied by emission of the visible light.  It should be    contrasted with   meteors and meteorites which are normally characterized by  strong emission in the visible frequency bands through sputtering and ablation \cite{Silber-review, Silber-optical-1, Silber-optical-2} and routinely observed by all-sky cameras.

\end{document}